\documentclass[aps,prapplied,reprint,superscriptaddress,floatfix]{revtex4-2} 

\usepackage[T1]{fontenc}
\usepackage{times}

\usepackage{graphicx}
\usepackage{booktabs}

\usepackage{amsfonts, amsmath,amssymb, bm, graphicx}
\usepackage{siunitx}
\DeclareSIUnit{\belmilliwatt}{Bm}
\DeclareSIUnit{\dBm}{\deci\belmilliwatt}

\usepackage{amssymb}
\usepackage[english]{babel}
\usepackage{lipsum}

\newcommand{\figref}[2]{\hyperref[#1]{\ref{#1}(#2)}}
\newcommand{\figrefsub}[3]{\hyperref[#1]{\ref{#1}(#2)#3}}

\usepackage{physics}
\usepackage{lipsum}
\usepackage{changes}
\usepackage[a]{esvect}
\usepackage[colorlinks=true,allcolors=blue]{hyperref}
\usepackage{footmisc}

\usepackage{upgreek}

\usepackage{soul}

\makeatletter
\let\ORIbbl@fixname\bbl@fixname
\def\bbl@fixname#1{%
  \@ifundefined{languagealias@\expandafter\string#1}
    {\ORIbbl@fixname#1}
    {\edef\languagename{\@nameuse{languagealias@#1}}}%
}
\newcommand{\definelanguagealias}[2]{%
  \@namedef{languagealias@#1}{#2}%
}
\makeatother

\definelanguagealias{en}{english}

\begin{document}

\title{Benchmarking a magnon-scattering reservoir with modal and temporal multiplexing}

\author{Christopher Heins}
\affiliation{Helmholtz-Zentrum Dresden--Rossendorf, Institut f\"ur Ionenstrahlphysik und Materialforschung, D-01328 Dresden, Germany}
\affiliation{Fakultät Physik, Technische Universität Dresden, D-01062 Dresden, Germany}

\author{Joo-Von Kim}
\affiliation{Centre de Nanosciences et de Nanotechnologies, CNRS, Université Paris-Saclay,
Palaiseau, 91120, France.}

\author{Lukas K\"orber}
\affiliation{Helmholtz-Zentrum Dresden--Rossendorf, Institut f\"ur Ionenstrahlphysik und Materialforschung, D-01328 Dresden, Germany}
\affiliation{Fakultät Physik, Technische Universität Dresden, D-01062 Dresden, Germany}
\affiliation{Radboud University, Institute of Molecules and Materials, Heyendaalseweg 135, 6525 AJ Nijmegen, The Netherlands}

\author{Jürgen Fassbender}
\affiliation{Helmholtz-Zentrum Dresden--Rossendorf, Institut f\"ur Ionenstrahlphysik und Materialforschung, D-01328 Dresden, Germany}
\affiliation{Fakultät Physik, Technische Universität Dresden, D-01062 Dresden, Germany}

\author{Helmut Schultheiss}\email{h.schultheiss@hzdr.de}
\affiliation{Helmholtz-Zentrum Dresden--Rossendorf, Institut f\"ur Ionenstrahlphysik und Materialforschung, D-01328 Dresden, Germany}

\author{Katrin Schultheiss}\email{k.schultheiss@hzdr.de}
\affiliation{Helmholtz-Zentrum Dresden--Rossendorf, Institut f\"ur Ionenstrahlphysik und Materialforschung, D-01328 Dresden, Germany}

\date{\today}

\begin{abstract}

Physical reservoir computing has emerged as a powerful framework for exploiting the inherent nonlinear dynamics of physical systems to perform computational tasks. Recently, we presented the magnon-scattering reservoir, whose internal nodes are given by the fundamental wave-like excitations of ferromagnets called magnons. These excitations can be geometrically-quantized and, in response to an external stimulus, show transient nonlinear scattering dynamics that can be harnessed to perform memory and nonlinear transformation tasks. Here, we test a magnon-scattering reservoir in a single magnetic disk in the vortex state towards two key performance indicators for physical reservoir computing, the short-term memory and parity-check tasks. Using time-resolved Brillouin light scattering microscopy, we measure the evolution of the reservoir's spectral response to an input sequence consisting of random binary inputs encoded in microwave pulses with two distinct frequencies. Two different output spaces of the reservoir are defined, one based on the time-averaged frequency spectra and another based on temporal multiplexing. Our results demonstrate that the memory and nonlinear transformation capability do not depend on the chosen read-out scheme as long as the dimension of the output space is large enough to capture all nonlinear features provided by the magnon-magnon interactions. This further shows that solely the nonlinear magnons in the physical system, not the read-out, determine the reservoir's capacity.

\end{abstract}

\maketitle

\section{Introduction}

Physical reservoir computing (PRC)~\cite{tanakaRecentAdvancesPhysical2019,nakajimaPhysicalReservoirComputing2020,liangPhysicalReservoirComputing2024} is a computational paradigm that exploits the natural dynamics of complex, nonlinear physical systems to process information. Originally inspired by recurrent neural networks, reservoir computing (RC)~\cite{jaegerHarnessingNonlinearityPredicting2004,10.1162/089976602760407955,verstraetenExperimentalUnificationReservoir2007,nakajimaReservoirComputingTheory2021} simplifies training by keeping a fixed set of randomly connected neurons — the reservoir — and only optimizing the output layer, rather than the entire network. This architecture has proven particularly effective in temporal tasks like pattern recognition, signal processing, and chaotic time-series prediction. In PRC, the reservoir is not an abstract neural network but an actual physical system with intrinsic nonlinear dynamics which map a complex input into a high dimensional, linearly separable output space. Limiting the problem to a simple linear classification, without the need for complex control over internal states, makes PRC a promising alternative for tasks requiring fast and energy-efficient information processing.

In recent years, various  substrates were suggested for physical reservoirs, such as optical fibers~\cite{duportAllopticalReservoirComputing2012,vandersandeAdvancesPhotonicReservoir2017,chemboMachineLearningBased2020,yanEmergingOpportunitiesChallenges2024}, memristive circuits~\cite{duReservoirComputingUsing2017,moonTemporalDataClassification2019,zhongDynamicMemristorbasedReservoir2021,zhongMemristorbasedAnalogueReservoir2022},  spintronic devices~\cite{kanaoReservoirComputingSpinTorque2019,tsunegiPhysicalReservoirComputing2019,markovicReservoirComputingFrequency2019,yamaguchi2020step,taniguchiSpintronicReservoirComputing2022,akashiCoupledSpintronicsNeuromorphic2022}, arrays of nano-magnetic elements~\cite{gartsideReconfigurableTrainingReservoir2022,allwoodPerspectivePhysicalReservoir2023,jensenClockedDynamicsArtificial2024}, skyrmions~\cite{prychynenkoMagneticSkyrmionNonlinear2018,pinnaReservoirComputingRandom2020,sunExperimentalDemonstrationSkyrmionenhanced2023}, and magnons~\cite{nakaneReservoirComputingSpin2018,pappCharacterizationNonlinearSpinwave2021,leeReservoirComputingSpin2022,nakanePerformanceEnhancementSpinWaveBased2023,namikiExperimentalDemonstrationHighPerformance2023,Nagase2024,namikiFastPhysicalReservoir2024}.
In our work, we study the computational capabilities of a magnon-scattering reservoir~\cite{korber2023pattern}. This realization relies on the nonlinear interactions of quantized magnons, the collective excitations in a magnetically ordered system. In particular, we evaluate the short-term memory (STM) and parity check (PC) tasks, both fundamental benchmarks for assessing the memory and nonlinear transformation capabilities of a computational system.
The STM task evaluates a system’s ability to maintain temporal information over short durations. This is critical for any temporal processing model, as many real-world signals involve retaining information over sequential time steps. 
The PC task tests the system's ability to perform nonlinear transformations, as determining parity involves identifying and processing specific nonlinear relationships among sequential inputs. 
Both tasks can be done on the same input sequence, with the difference of separately trained output layers that map the response of the magnon-scattering reservoir to the respective transformed target.

In our experiments, the temporal evolution of the magnon reservoir's spectral response is measured by means of time-resolved Brillouin light scattering microscopy. 
To characterize and read-out the magnon-scattering reservoir, we compare two distinct output spaces of the reservoir: one based on time-averaged frequency spectra of the magnon system, another defined by the temporal evolution of the individual magnon modes.

\section{Methods}

\subsection{Sample preparation}

\begin{figure}[]
    \centering
    \includegraphics{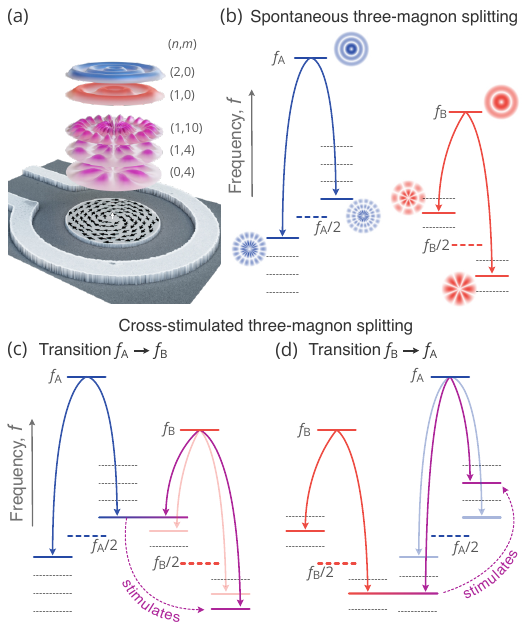}
    \caption{(a) Magnon-scattering reservoir consisting of a \SI{50}{\nano\meter} thick,  \SI{5.1}{\micro\meter} diameter NiFe disk in the vortex state placed inside an omega-shaped antenna, with schematic intensity profiles of different magnon modes $(n,m)$. (b) Two radial magnon modes $f_\text{A}$ and $f_\text{B}$ spontaneously split into two magnons with $m \neq 0$ and frequencies equally spaced around half their respective excitation frequencies. (c),(d) The pulsed excitation of $f_\text{A}$ and $f_\text{B}$ with a temporal overlap results in cross-stimulation of the two channels, where the split modes of the later pulse depend on the temporal order of the two inputs.}
    \label{fig:1}
\end{figure}

The magnon-scattering reservoir is based on a single magnetic disk with a diameter of \SI{5.1}{\micro\meter} which is patterned from a Ti(\SI{2}{\nano\meter})/Ni$_{81}$Fe$_{19}$(\SI{50}{\nano\meter})/Ti(\SI{5}{\nano\meter}) film deposited on a SiO$_2$ substrate using electron beam lithography, electron beam evaporation and subsequent lift off. To excite magnetization dynamics, we pattern an $\Omega$-shaped antenna around the disk from a Ti(\SI{2}{\nano\meter})/Au(\SI{200}{\nano\meter}) film, also using electron beam lithography, electron beam evaporation and lift off. The inner and outer diameter of the antenna are \SI{8.3}{\micro\meter} and \SI{11.1}{\micro\meter}, respectively. An illustration of the sample, which is based on a scanning electron microscopy (SEM) image, is shown in Fig.~\figref{fig:1}{a}.

As depicted on the SEM image, the ground state in such a disk is a magnetic vortex with the magnetic moments (black arrows) curling in-plane around the vortex core, a narrow region in the disk's center where the magnetic moments align perpendicularly to the film plane (white arrow). The magnon eigenmodes in this system have well-defined frequencies and are described by their mode numbers ($n, m$), with $n$ ($m$) counting the number of nodes in radial (azimuthal) direction, respectively. A few examples of mode profiles are depicted in Fig.~\figref{fig:1}{a}. By applying a microwave current to the $\Omega$-shaped antenna, the symmetry of the homogeneous out-of-plane Oersted field solely allows for the excitation of pure radial modes with $m=0$.

\subsection{Nonlinear three-magnon splitting}

If the excitation power crosses a certain threshold, nonlinear three-magnon splitting sets in~\cite{schultheiss_excitation_2019,korberNonlocalStimulationThreeMagnon2020}, as is sketched in Fig.~\figref{fig:1}{b}. The initially excited magnon mode ($n_\textrm{A}, 0$) at frequency $f_\textrm{A}$ spontaneously splits into two secondary magnon modes under conservation of energy $f_\textrm{A}=f_\textrm{A+}+f_\textrm{A-}$ and angular momentum $m_\textrm{A+}=-m_\textrm{A-}$. Inside a single disk, many different channels for this splitting exist, as indicated in Fig.~\figref{fig:1}{b} for excitation frequencies $f_\textrm{A}$ and $f_\textrm{B}$. 

These different splitting channels can mutually influence each other via cross-stimulated three-magnon scattering.~\cite{korber2023pattern} During transient dynamics, this influence between different channels can be nonreciprocal.
Figure~\figref{fig:1}{c, d} depicts the simplified case with two magnon modes excited by ns-long microwave pulses.
The spontaneous splitting of frequency $f_\text{A}$ forces $f_\text{B}$ into a specific pair of split magnons, which deviate from its spontaneous split modes [Fig.~\figref{fig:1}{c}].
If the temporal order of the two pulses is reversed [Fig.~\figref{fig:1}{d}], $f_\text{B}$ splits spontaneously and forces $f_\text{A}$ into different split modes during the transient dynamics. Therefore, it is possible to distinguish the temporal order of the input signals by only measuring the nonlinearly excited split modes. This nonlinear interaction of quantized magnon modes in reciprocal space underpins the magnon-scattering reservoir.

\begin{figure*}[]
    \centering
    \includegraphics{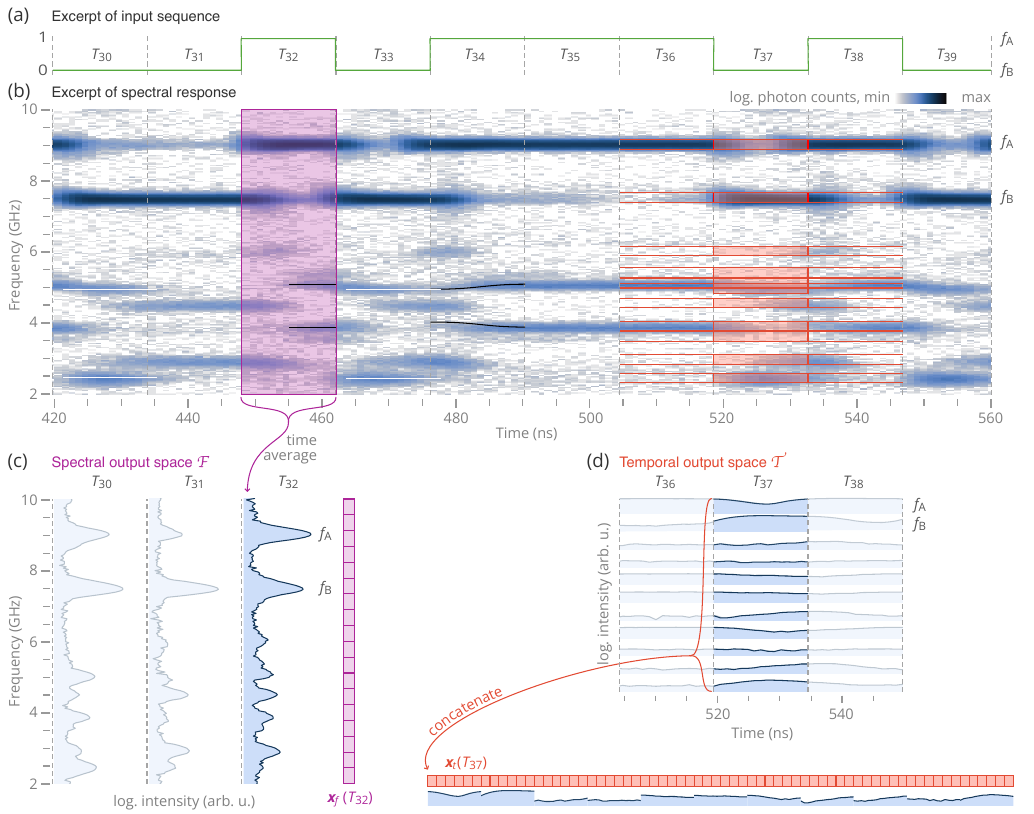}
    \caption{(a) Excerpt of a binary input sequence "$0010111010$" which is encoded by alternating microwave pulses, each $\SI{14}{\ns}$ long and resembling one time step $T_j$. "$1$" ("$0$") is represented by the frequency $f_\text{A}$ $(f_\text{B})$, respectively. (b) Excerpt of the time-resolved µBLS data measured on the magnon-scattering reservoir with the recorded magnon intensity color-coded. The colored boxes highlight areas of the spectrum that are used to evaluate the system's performance based on two distinct output schemes. (c) On the one hand, the spectral response is integrated for each time step $T_j$, and the binned intensity is used to construct output space $\mathcal{F}$. (d) On the other hand, the time evolution of all involved magnon modes is tracked individually and then concatenated to generate output space $\mathcal{T}$.}
    \label{fig:2}
\end{figure*}

\subsection{Signal generation}

To perform the STM and PC task experimentally, we generate a sequence of 2000 random binary inputs $({\textit{u}} \in [0,1])$, an excerpt of which is shown in Fig.~\figref{fig:2}{a}. This input is then transformed into 2000 radiofrequency (rf) pulses, each with a nominal length of \SI{14}{\nano\second} and representing an individual time step $T_j$ with $0 \leq j\leq 2000$. The pulse duration is chosen to be as short as possible to have the pulse transitions in a time window, where the dynamics are not yet in steady state. For the used microwave equipment this corresponds to \SI{14}{\nano\second}. The binary input "$1$" corresponds to the third radial magnon mode ($2, 0$) excited at frequency $f_\textrm{A}=\SI{8.9}{\giga\hertz}$, the input "$0$" corresponds to the second radial mode ($1, 0$) at frequency $f_\textrm{B}=\SI{7.4}{\giga\hertz}$. The excitation powers $P_\textrm{A}=\SI{20}{\dBm}$ and $P_\textrm{B}=\SI{24}{\dBm}$ are strong enough to reach the nonlinear regime spontaneously for each of the targeted modes A and B. Due to the higher threshold for $f_\textrm{B}$, the power is larger for that frequency to achieve similar intensities of the split modes. 

The rf pulses are generated by two separate microwave sources set to the pulse mode at fixed frequencies $f_\textrm{A}$ and $f_\textrm{B}$, respectively. In order to synchronize the two sources, a pattern generator (Pulse\-streamer by Swabian Instruments) is used to create the series of pulses gating the rf-sources. The two generated microwave signals are combined and fed into the $\Omega$-shaped antenna using picoprobes. 

\subsection{Time-resolved Brillouin light scattering microscopy}

The response of the magnon-scattering reservoir to the input sequence is probed as a function of frequency and time by time-resolved micro-focused Brillouin light scattering spectroscopy (TR-µBLS) ~\cite{sebastianMicrofocusedBrillouinLight2015}.
A monochromatic (532 nm) continuous-wave laser is focused onto the sample surface using a 100x microscope lens with a high numerical aperture (0.75), allowing us to reach a spatial resolution of about \SI{300}{\nano\meter}. The backscattered light is sent into a Tandem Fabry-P\'{e}rot interferometer (TFPI) ~\cite{mockConstructionPerformanceBrillouin1987} to measure the frequency shift caused by inelastic scattering of photons and magnons. 
Control signals that encode the current state of the interferometer, signals of the photon counter and a clock signal from the pattern generator are acquired continuously by a time-to-digital converter (Timetagger 20 by Swabian Instruments) with a temporal resolution of \SI{1}{\nano\second}. From these data, the temporal evolution of the magnon spectra with respect to the stroboscopic rf excitation is reconstructed. 

During the experiments, the investigated structure is imaged using a red LED and a CCD camera. Displacements and drifts of the sample are tracked by an image recognition algorithm and compensated by the positioning system (XMS linear stages by Newport).
To capture magnon modes with different spatial distributions, we integrate the signal over 12 positions covering one-quarter of the disk, four in azimuthal and three in radial direction. 
All measurements are conducted at room temperature.

\subsection{Micromagnetics simulations}

To support the experimental work, we also performed micromagnetic simulations of the magnon-scattering reservoir response to the same sequence of rf pulses. We used the MuMax3 code which performs numerical time-integration of the Landau-Lifshitz-Gilbert equation using the finite difference method~\cite{vansteenkiste_design_2014}. We modeled the \SI{5}{\micro\meter} diameter, \SI{50}{\nano\meter} thick permalloy disk using $512 \times 512 \times 1$ cells, with an exchange constant of $A = \SI{13}{\pico\joule/\meter}$, a saturation magnetization of $M_\text{S} = \SI{810}{\kilo\ampere/\meter}$, a Gilbert damping constant of $\alpha = 0.008$, and a gyromagnetic constant of $\gamma/2\pi = \SI{29.6}{\giga\hertz/\tesla}$. Finite temperature effects are accounted for by adding a stochastic term to the effective field, where we assume $T = \SI{300}{\kelvin}$. The rf field excitation is modeled with a spatially-uniform perpendicular field, $\mathbf{b}_\mathrm{rf} = b_i \sin(2 \pi f_i t) \hat{\mathbf{z}}$, where $b_\text{A}= \SI{3.0}{\milli\tesla}$ for $f_\text{A}  = \SI{8.9}{\giga\hertz}$ and $b_\text{B} = \SI{3.5}{\milli\tesla}$ for $f_\text{B}  = \SI{7.4}{\giga\hertz}$. 

Following the approach outlined in Ref.~~\citenum{korber2023pattern}, we estimate the power spectra of magnon excitations using a coarse-grained approach. We subdivide the simulation geometry into a triangle mesh and record the spatial average of the magnetization over each mesh region as a function of time, with a time step of \SI{20}{\pico\second}. The total power spectrum is then computed by summing the power spectra of these mesh regions, obtained using a discrete Fourier transform of the fluctuating magnetization over an interval corresponding to the pulse length of \SI{14}{\nano\second}. This results in a sequence of 2000 spectra, corresponding to a random input sequence of 2000 rf pulses with frequencies $f_\text{A}$ and $f_\text{B}$. To mimic the experiment as closely as possible, which involves multiple repetitions of the same pulse sequence, we repeat this simulation 100 times, but with a different realisation of the stochastic thermal field. The average power spectrum over these 100 runs is subsequently used for computing the STM and PC metrics.

\section{Results and discussion}

An excerpt of the magnon response is shown in Fig.~\figref{fig:2}{b}, with the recorded magnon intensity color-coded on a logarithmic scale. Due to the long measurement duration of \SI{28.1}{\micro\second}, the full input sequence and the complete time-resolved BLS spectrum are not shown, but available in Ref.~~\citenum{heins_christopher_2025_3558}. The gray dashed lines separate the \SI{14}{\nano\second} long time steps $T_j$. The directly excited modes at $f_\textrm{A}$ and $f_\textrm{B}$ yield the most intense response. Their duration is slightly longer than the supplied input pulses due to the finite magnon lifetime.

Besides the directly excited modes, one can see a multitude of secondary modes in the frequency range from \SIrange{2}{7}{\giga\hertz}, i.e. around half the frequencies of the initially excited modes. As explained in the previous section, the secondary modes are subject to continuous cross-stimulation from one another.
However, in addition to the simple cross-stimulation of two consecutive pulses depicted Fig.~\figref{fig:1}{c,d}, the split modes in a longer input sequence also depend on excitation pulses further in the past.
An example of this long-term manipulation can be seen in the TR-µBLS spectrum [Fig.~\figref{fig:2}{b}] when comparing the transitions $T_{31}\to T_{32}$ and $T_{33} \to T_{34}$. Both cases show the transition $f_\text{B} \to f_\text{A}$, but the intensities of the split modes are clearly different due to the long-term history. 

At the beginning of time step $T_{33}$, $f_\text{B}=\SI{7.4}{\giga\hertz}$ is forced to split into modes at \SI{2.5}{\giga\hertz} and \SI{4.9}{\giga\hertz} [horizontal white lines in $T_{33}$ of Fig.~\figref{fig:2}{b}]. For the transition $T_{33} \to T_{34}$, the already existing population of the mode at \SI{4.9}{\giga\hertz} stimulates the splitting of $f_\text{A}=\SI{8.9}{\giga\hertz}$ into modes at \SI{3.8}{\giga\hertz} and \SI{5.1}{\giga\hertz} directly at the beginning of $T_{34}$, even though slightly off-resonant [black lines in $T_{34}$ of Fig.~\figref{fig:2}{b}]. However, the initial mode populations are different for the transition $T_{31} \to T_{32}$. At the beginning of $T_{32}$, the mode at \SI{4.9}{\giga\hertz} is not populated from the previous time step anymore due to the limited lifetime of the mode at \SI{2.5}{\giga\hertz}. Therefore, it takes almost \SI{7}{\nano\second} for the splitting of $f_\text{A}$ to build up in $T_{32}$ [horizontal black lines in Fig.~\figref{fig:2}{b}]. 
This gives one example of how the splitting of the directly excited modes is influenced by time steps further in the past. 

To analyze the memory and nonlinear capabilities of the magnon-scattering reservoir more quantitatively, we use two different output schemes for our analysis.
The first output space $\mathcal{F}$ is constructed following modal multiplexing introduced in Ref.~~\citenum{korber2023pattern}. As is shown in Fig.~\figref{fig:2}{c} for one time step $T_{32}$, the spectral response is integrated over time within each \SI{14}{\nano\second} long pulse. Then the integrated spectra are separated into $N_f$ different frequency bins, in which the detected intensities are averaged. This is done for each time step $T_j$. The resulting vectors $\textbf{\textit{x}}_f(T_j)$ form a matrix $\textbf{\textit{X}}_f$ representing the output space $\mathcal{F}$.

Alternatively, a more common approach for reservoir computing in the time domain is used, as illustrated in Fig.~\figref{fig:2}{d}. Here, the intensities of the ten most intense magnon modes are tracked by integrating a \SI{300}{\mega\hertz} window around the center frequency of the respective mode. Then, these time traces for the different modes are appended to form one long time-stream, which is divided into a number $N_t$ of time steps. The resulting vectors $\textbf{\textit{x}}_t(T_j)$ for different time steps are concatenated to create the output matrix $\textbf{\textit{X}}_t$. This defines the temporal output space $\mathcal{T}$.

Both output spaces, $\mathcal{F}$ and $\mathcal{T}$, are extracted for different frequency bin sizes and for varying resolutions of the time traces, respectively. This allows for a direct comparison between output spaces with similar dimensions $N_f\sim N_t$, which will be explained in more detail below.

\begin{figure}[]
    \centering
    \includegraphics{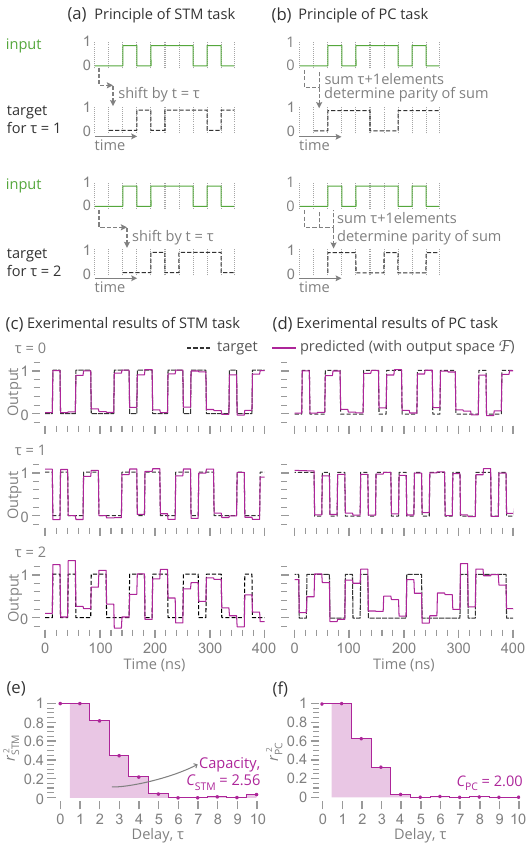}
    \caption{ Principle of (a) the short-term memory (STM) task and (b) the parity check (PC) task plotting the target outputs for the first two delays $\tau=1,2$. For the STM task, the input is shifted forward in time by $\tau$. The PC task determines the parity of the sum over the previous $(\tau +1)$ time steps. (c),(d) With the spectral output space $\mathcal{F}$, the output is predicted for different delays $\tau$ (continuous lines) and is compared to the target values (dashed lines). (e),(f)  Correlation between the target and predicted output as a function of delay $\tau$ for (e) the STM task and (f) the PC task. The colored areas yield the respective capacities $C$ for each task.}
    \label{fig:3}
\end{figure}

Following calculations of the memory capacity in Refs.~~\citenum{jaegerShortTermMemory2001,daleSubstrateindependentFrameworkCharacterize2019}, 
the target vectors $\hat{\textbf{\textit{y}}}_{\textrm{STM}}(\tau)$  for the STM task are constructed by delaying the original binary input sequence by $\tau$ time steps into the future, as illustrated for $\tau=1,2$ in Fig.~\figref{fig:3}{a}:
\begin{equation}
    \hat{{\textit{y}}}_{\textrm{STM}}(T_{j},\tau) = {\textit{u}}(T_{j-\tau})
\end{equation}
This results in the current output $\textbf{\textit{x}}(T_{j})$ at time step $T_j$, being used to recall the preceding input $\textit{u}(T_{j-\tau})$ at $\tau$ steps in the past.

For the PC task, the target vectors $\hat{\textbf{\textit{y}}}_{\textrm{PC}}(\tau)$ are expressed as the parity of the sum of the last ($\tau+1$) inputs of the binary sequence $\textbf{\textit{u}}$, as shown for $\tau=1,2$ in Fig.~\figref{fig:3}{b}. This is equivalent to $\tau$ consecutive XOR ($\oplus$) operations on the input sequence:
\begin{equation}
    \hat{{\textit{y}}}_{\textrm{PC}}(T_{j},\tau) = \textit{u}(T_{j}) \oplus \textit{u}(T_{j-1}) \oplus ... \oplus \textit{u}(T_{j-\tau})  
\end{equation}

The reservoir's output layer is then trained to map the experimentally acquired output matrix $\textbf{\textit{X}}$ to the previously described target functions with \textbf{\textit{W}} being the matrix of the individual output weights. For each task, the weight matrix is trained separately: 
\begin{equation}
\textbf{\textit{W}}_\textrm{STM} \textbf{\textit{X}} = \hat{\textbf{\textit{Y}}}_{\textrm{STM}}
\end{equation}
\begin{equation}
\textbf{\textit{W}}_\textrm{PC} \textbf{\textit{X}} = \hat{\textbf{\textit{Y}}}_{\textrm{PC}}
\end{equation}
To find the matrices $\textbf{\textit{W}}$, the output is split into a training data set containing 1550 pulses and a set of 400 testing pulses. The first ten pulses are discarded to let the system settle in the beginning.
With the pseudo inverse of the training pulse output, one can construct the weight matrix for each delay $\tau$ of the STM and the PC task: 
\begin{equation}
 \textbf{\textit{W}}_\textrm{STM}= \hat{\textbf{\textit{Y}}}_{\textrm{STM}}\textbf{\textit{X}}^{-1}
\end{equation}
\begin{equation}
 \textbf{\textit{W}}_\textrm{PC}= \hat{\textbf{\textit{Y}}}_{\textrm{PC}}\textbf{\textit{X}}^{-1}
\end{equation}

\begin{figure*}[]
    \centering
    \includegraphics{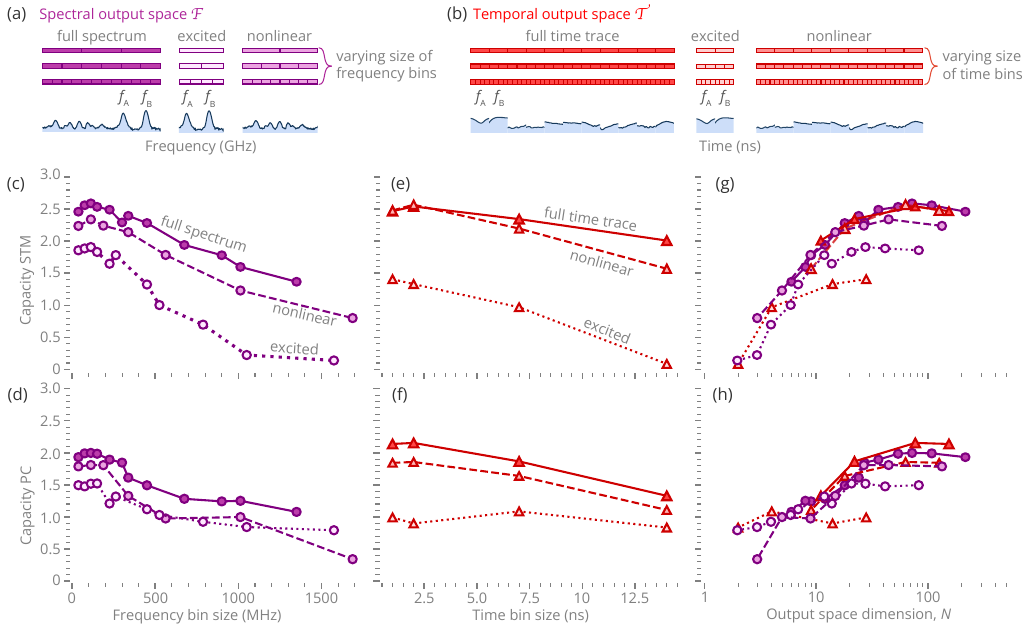}
    \caption{(a) Distinction between full spectrum, directly excited modes and nonlinear split modes, which are used to determine reservoir metrics based on the spectral output space $\mathcal{F}$. Colored bars with different sizes indicate varying sizes of the frequency bins. (b) Distinction between full time trace, directly excited modes and nonlinear split modes, which are used to determine reservoir metrics based on the temporal output space $\mathcal{T}$. Colored bars with different sizes indicate varying sizes of the time bins. (c)-(f) Comparison of STM capacity (upper row) and PC capacity (bottom row) as a function of the bin size and based on the different output spaces. (g),(h) Capacities plotted as a function of output space dimension $N$ independent from its type.}
    \label{fig:4}
\end{figure*}

In Fig.~\figref{fig:3}{c,d}, the calculated weight matrices are used to predict the output for the STM and PC tasks based on the test data of the spectral output space $\mathcal{F}$. Note that we plot the results as a function of absolute time $t$, with \SI{14}{\nano\second} long rf pulses corresponding to one time step $T_j$ of the binary input sequence and the axis starting at the first pulse after the training data. It can be seen that the reconstruction works reasonably well for the first few delays $\tau$, which confirms that the magnon-scattering reservoir exhibits short-term memory and nonlinear transformation capabilities. 

This is further quantified by calculating the square of the correlation coefficient between the predicted output 
$\textbf{\textit{y}}_\textrm{out}$ and the expected target output $\hat{\textbf{\textit{y}}}_\textrm{out}$:
\begin{equation}
    r^{2}(\tau)=\frac{\text{cov}({\textbf{\textit{y}}}_\textrm{out}(\tau),\hat{\textbf{\textit{y}}}_\textrm{out}(\tau))^2}{\text{cov}({\textbf{\textit{y}}}_\textrm{out}(\tau),{\textbf{\textit{y}}}_\textrm{out}(\tau))\times\text{cov}(\hat{\textbf{\textit{y}}}_\textrm{out}(\tau),\hat{\textbf{\textit{y}}}_\textrm{out}(\tau))}
\end{equation}
where $\textrm{cov}(\textbf{\textit{x}},\textbf{\textit{y}})$ is the covariance between vectors $\textbf{\textit{x}}$ and $\textbf{\textit{y}}$. A value of 1 for the correlation coefficient corresponds to a perfect prediction, while values close to 0 mean the prediction is random.
Figure~\figref{fig:3}{e,f} shows the resulting squared correlations up to $\tau=10$ for the STM and PC tasks, respectively. Note that at zero delay ($\tau = 0$), the direct input is simply reconstructed and, hence, is neglected for the further evaluation. Therefore, the capacity of the reservoir is defined by the area below the remaining squared correlation function shaded in purple: 
\begin{equation}
    C_\textrm{}=\sum_{\tau=1}^{10}r_\textrm{}^2(\tau)
\end{equation}

To systematically assess the impact of different types of output spaces, we repeat our analysis for the spectral output space $\mathcal{F}$ as well as for the temporal output space $\mathcal{T}$. In addition, we further distinguish different data ranges for our analysis to better understand the origin of the reservoir's memory and computing capabilities, as illustrated in Fig.~\figref{fig:4}{a,b}. 

For the spectral output space $\mathcal{F}$, we consider three different frequency ranges [Fig.~\figref{fig:4}{a}]. First, we begin with the full spectrum from \SIrange{2}{10}{\giga\hertz} to evaluate the reservoir's full potential.
Second, we limit our analysis to the frequency range from \SIrange{7}{10}{\giga\hertz}. This range is closest to a simple linear transformation of the input because it only contains the directly excited modes and no nonlinear split modes. Nonetheless, these excited modes are also influenced by nonlinear intensity changes and frequency shifts.
Third, we choose a frequency window from \SIrange{2}{7}{\giga\hertz} focusing on the nonlinear split modes only. Magnons in this range are purely excited by nonlinear three-magnon splitting and, therefore, nicely capture the reservoir's nonlinear properties.
For the temporal output space $\mathcal{T}$, we make a similar distinction [Fig.~\figref{fig:4}{b}], first looking at the full time trace and then separating the directly excited modes from the nonlinear split modes. Additionally, we vary the bin sizes for the spectral and temporal binning to study the influence of the output space dimension and allow for a more direct comparison between the different output spaces with similar dimensions of the weight matrices.

Figure~\figref{fig:4}{c,d} summarizes the capacities for the STM and PC tasks which are analyzed based on the spectral output space $\mathcal{F}$. The memory capacity of the magnon-scattering reservoir is largest when combining all modes and reaches its maximum of $C_{\textrm{STM}}=2.59$ for a bin size of \SI{112.5}{\mega\hertz} [Fig.~\figref{fig:4}{c}]. Comparing the nonlinear and excited modes indicates that the largest contribution to this memory capacity originates from the nonlinear split modes. However, even the directly excited modes contribute as evidenced by the fact that their spectra also contain nonlinear contributions. Up to a frequency bin size of \SI{200}{\mega\hertz}, the memory capacity only changes marginally.
Further increasing the frequency bins, however, results in a significant drop in the memory capacity, which is even more pronounced in the directly excited modes than in the nonlinear split modes. For larger bin sizes above \SI{1}{\giga\hertz}, the memory is completely lost for the excited modes. 

The PC task shows a similar behavior for smaller bin sizes [Fig.~\figref{fig:4}{d}]. When analyzing the full spectrum, the nonlinear transformation capability reaches a maximum of $C_{\textrm{PC}}=2.0$ for a bin size of \SI{112.5}{\mega\hertz} and is rather constant up to \SI{300}{\mega\hertz}. The nonlinear modes contribute most to the nonlinear transformation capabilities of the magnon-scattering reservoir. Nonetheless, the performance is improved by including the directly excited modes as well. Interestingly, for bin sizes larger than \SI{300}{\mega\hertz}, the nonlinear capacity is similar when analyzing the excited modes and the nonlinear modes separately. 

For the temporal output space $\mathcal{T}$, the directly excited modes only contain the time evolution of a \SI{300}{\mega\hertz} integrated window around \SI{7.4}{\giga\hertz} and \SI{8.9}{\giga\hertz}. The chosen scattered modes are shown in Fig.~\figref{fig:1}{d} and correspond to the nine most intense split modes. For the smallest time bin size, the memory capacity of the nonlinear modes [Fig.~\figref{fig:4}{e}] shows very similar results to the spectral output space $\mathcal{T}$. Also, it can be seen that for the memory, the nonlinear modes contain all relevant information, and adding the excited modes does not further increase the maximum value of $C_{\textrm{STM}}=2.54$ at \SI{2}{\nano\second} bin size. 
Interestingly, this is not the case for the PC task. Here, the capacity improves when analyzing the full-time trace compared to only evaluating the nonlinear response. This is surprising since the system's nonlinearity in the form of three-magnon splitting should be more relevant for the PC task than for the STM task. Overall, the capacity for the parity check reaches a slightly higher maximum capacity $C_{\textrm{PC}}=2.15$ at a bin size of \SI{2}{\nano\second} than for the spectral output space $\mathcal{F}$.

In general, comparing the spectral and temporal output spaces shows that the scattered modes are more important for the metrics of the magnon-scattering reservoir but that the directly excited frequencies also contain valuable information, whereby their inclusion in the output space improves the overall capacities. 

Changing the bin sizes has direct impact on the number of output features. Therefore, the different output spaces are compared regarding their dimension $N$ in Fig.~\figref{fig:4}{g,h}. This shows that the directly excited modes perform worse when considered alone for the same amount of features. However, combining both directly excited and nonlinear modes yields better results even with a similar amount of output variables. Overall, the performance depends not on which output space was chosen but on the number of features when considering a combination of excited and nonlinear modes. In addition, one can see that too many features degrade the capacity due to too many training weights, resulting in overfitting.

\begin{figure}
	\includegraphics{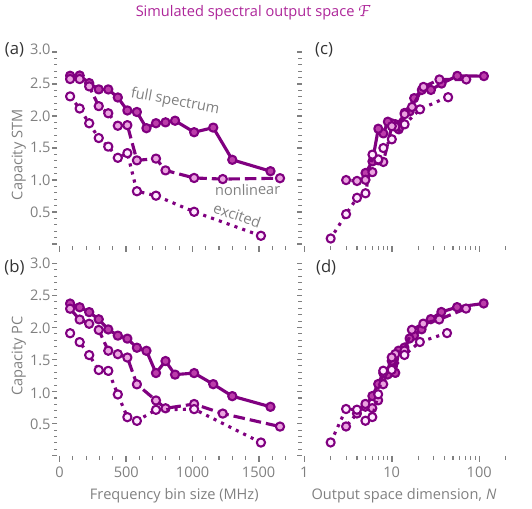}
	\caption{Reservoir computing metrics obtained from micromagnetic simulations. STM (a),(c) and PC (b),(d) capacities as a function of bin size (a),(b) and output space dimension $N$ (c),(d).}
	\label{fig:simulations}
\end{figure}

Similar results are obtained in micromagnetic simulations for the STM and PC capacities using the output space $\mathcal{F}$, as shown in Fig.~\ref{fig:simulations}.
As in the experiments, we observe the best performance when the full spectrum is used to construct the output space vector, which slightly outperforms the case when only the nonlinear excitations are used. Overall, the capacities for both the STM and PC metrics are very close to the experimental results with maximum values of $C_\textrm{STM}=2.57$ and $C_\textrm{PC}=2.37$. They follow a similar decrease as a function of the bin size and increase with the output space dimension. This also indicates that the experimental values stem from the magnon-scattering reservoir itself and are unrelated to possible nonlinearities of the microwave circuitry.

\section{Conclusion}

In order to benchmark the memory and nonlinear transformation potential of a magnon-scattering reservoir comprising a single micrometer-sized magnetic disk in the vortex state, we have performed the short-term memory and the parity check tasks, which are two established metrics for physical reservoir computing devices. For this, a random binary sequence was encoded into two oscillation frequencies in the GHz regime and used to excite dynamics in the magnon reservoir using a microwave antenna. The resulting highly complex nonlinear scattering of the quantized magnon modes, which gives rise to the reservoir computing capabilities, was subsequently measured using time-resolved Brillouin light scattering microscopy and used to define two distinct types of reservoir output spaces. In addition to the modal output space $\mathcal{F}$, given by the frequency spectra averaged over the time intervals of the input sequence, a temporal output space $\mathcal{T}$ was used, which was constructed by directly tracking the temporal evolution of the ten most dominant modes, averaged along a small frequency window around the central frequency of the respective mode.
The capacity reached maximum values of $C_{\textrm{STM}}=2.59$ for the short-term memory task and $C_{\textrm{PC}}=2.15$ for the parity-check task which is comparable to other physical substrates~\cite{tsunegiPhysicalReservoirComputing2019,yamaguchi2020step,Nagase2024}.

Importantly, for the optimal output dimension, the two output spaces $\mathcal{F}$ and $\mathcal{T}$ yield comparable capacities. On the one hand, this highlights that it is not necessary to track the full temporal evolution of the scattering reservoir to achieve the best performance. Instead, as long as the nonlinear magnon dynamics are sufficiently well resolved, it is equivalent to tracing the time-averaged spectra of the active modes in the system. On the other hand, the equivalence between the two output spaces clearly shows that neither read-out scheme changes the nonlinearity of the system in a relevant way. In light of the fact that in physical reservoir computing, it can be difficult to disentangle how much capacity is carried by the (possibly nonlinear) read-out scheme, this clearly shows that the obtained capacities in this work are dominated solely by the nonlinear magnon interactions.

\section*{Author declarations}

\subsection*{Conflict of Interest}
The authors have no conflicts of interest to disclose.

\subsection*{Author's contributions}

H.S., C.H., and K.S conceptualized the presented work. 
J.V.K., H.S., and K.S. acquired funding.
K.S. fabricated the sample.
C.H. conducted the experiments and analyzed the data.
J.V.K. performed and analyzed the micromagnetic simulations.
All authors discussed the results.
C.H., H.S., L.K., and K.S. visualized the results.
C.H., L.K. and K.S. wrote the original draft of the paper. 
All authors reviewed and edited the paper.


\section*{Acknowledgements}

This work received funding by the EU Research and Innovation Programme Horizon Europe under grant agreement no. 101070290 (NIMFEIA).
Support by the Nanofabrication Facilities Rossendorf (NanoFaRo) at the IBC is gratefully acknowledged.

\section*{Data availability}
The data that support the findings of this study are openly available in Ref. ~~\citenum{heins_christopher_2025_3558}. In this work, the authors used the scientific colour map \textit{oslo} from (https://www.fabiocrameri.ch/colourmaps/).

\bibliography{references_new.bib}

\begin{thebibliography}{43}%
\makeatletter
\providecommand \@ifxundefined [1]{%
 \@ifx{#1\undefined}
}%
\providecommand \@ifnum [1]{%
 \ifnum #1\expandafter \@firstoftwo
 \else \expandafter \@secondoftwo
 \fi
}%
\providecommand \@ifx [1]{%
 \ifx #1\expandafter \@firstoftwo
 \else \expandafter \@secondoftwo
 \fi
}%
\providecommand \natexlab [1]{#1}%
\providecommand \enquote  [1]{``#1''}%
\providecommand \bibnamefont  [1]{#1}%
\providecommand \bibfnamefont [1]{#1}%
\providecommand \citenamefont [1]{#1}%
\providecommand \href@noop [0]{\@secondoftwo}%
\providecommand \href [0]{\begingroup \@sanitize@url \@href}%
\providecommand \@href[1]{\@@startlink{#1}\@@href}%
\providecommand \@@href[1]{\endgroup#1\@@endlink}%
\providecommand \@sanitize@url [0]{\catcode `\\12\catcode `\$12\catcode
  `\&12\catcode `\#12\catcode `\^12\catcode `\_12\catcode `\%12\relax}%
\providecommand \@@startlink[1]{}%
\providecommand \@@endlink[0]{}%
\providecommand \url  [0]{\begingroup\@sanitize@url \@url }%
\providecommand \@url [1]{\endgroup\@href {#1}{\urlprefix }}%
\providecommand \urlprefix  [0]{URL }%
\providecommand \Eprint [0]{\href }%
\providecommand \doibase [0]{https://doi.org/}%
\providecommand \selectlanguage [0]{\@gobble}%
\providecommand \bibinfo  [0]{\@secondoftwo}%
\providecommand \bibfield  [0]{\@secondoftwo}%
\providecommand \translation [1]{[#1]}%
\providecommand \BibitemOpen [0]{}%
\providecommand \bibitemStop [0]{}%
\providecommand \bibitemNoStop [0]{.\EOS\space}%
\providecommand \EOS [0]{\spacefactor3000\relax}%
\providecommand \BibitemShut  [1]{\csname bibitem#1\endcsname}%
\let\auto@bib@innerbib\@empty
\bibitem [{\citenamefont {Tanaka}\ \emph {et~al.}(2019)\citenamefont {Tanaka},
  \citenamefont {Yamane}, \citenamefont {H{\'e}roux}, \citenamefont {Nakane},
  \citenamefont {Kanazawa}, \citenamefont {Takeda}, \citenamefont {Numata},
  \citenamefont {Nakano},\ and\ \citenamefont
  {Hirose}}]{tanakaRecentAdvancesPhysical2019}%
  \BibitemOpen
  \bibfield  {author} {\bibinfo {author} {\bibfnamefont {G.}~\bibnamefont
  {Tanaka}}, \bibinfo {author} {\bibfnamefont {T.}~\bibnamefont {Yamane}},
  \bibinfo {author} {\bibfnamefont {J.~B.}\ \bibnamefont {H{\'e}roux}},
  \bibinfo {author} {\bibfnamefont {R.}~\bibnamefont {Nakane}}, \bibinfo
  {author} {\bibfnamefont {N.}~\bibnamefont {Kanazawa}}, \bibinfo {author}
  {\bibfnamefont {S.}~\bibnamefont {Takeda}}, \bibinfo {author} {\bibfnamefont
  {H.}~\bibnamefont {Numata}}, \bibinfo {author} {\bibfnamefont
  {D.}~\bibnamefont {Nakano}},\ and\ \bibinfo {author} {\bibfnamefont
  {A.}~\bibnamefont {Hirose}},\ }\bibfield  {title} {\bibinfo {title} {Recent
  advances in physical reservoir computing: {{A}} review},\ }\href
  {https://doi.org/10.1016/j.neunet.2019.03.005} {\bibfield  {journal}
  {\bibinfo  {journal} {Neural Networks}\ }\textbf {\bibinfo {volume} {115}},\
  \bibinfo {pages} {100} (\bibinfo {year} {2019})}\BibitemShut {NoStop}%
\bibitem [{\citenamefont
  {Nakajima}(2020)}]{nakajimaPhysicalReservoirComputing2020}%
  \BibitemOpen
  \bibfield  {author} {\bibinfo {author} {\bibfnamefont {K.}~\bibnamefont
  {Nakajima}},\ }\bibfield  {title} {\bibinfo {title} {Physical reservoir
  computing---an introductory perspective},\ }\href
  {https://doi.org/10.35848/1347-4065/ab8d4f} {\bibfield  {journal} {\bibinfo
  {journal} {Japanese Journal of Applied Physics}\ }\textbf {\bibinfo {volume}
  {59}},\ \bibinfo {pages} {060501} (\bibinfo {year} {2020})}\BibitemShut
  {NoStop}%
\bibitem [{\citenamefont {Liang}\ \emph {et~al.}(2024)\citenamefont {Liang},
  \citenamefont {Tang}, \citenamefont {Zhong}, \citenamefont {Gao},
  \citenamefont {Qian},\ and\ \citenamefont
  {Wu}}]{liangPhysicalReservoirComputing2024}%
  \BibitemOpen
  \bibfield  {author} {\bibinfo {author} {\bibfnamefont {X.}~\bibnamefont
  {Liang}}, \bibinfo {author} {\bibfnamefont {J.}~\bibnamefont {Tang}},
  \bibinfo {author} {\bibfnamefont {Y.}~\bibnamefont {Zhong}}, \bibinfo
  {author} {\bibfnamefont {B.}~\bibnamefont {Gao}}, \bibinfo {author}
  {\bibfnamefont {H.}~\bibnamefont {Qian}},\ and\ \bibinfo {author}
  {\bibfnamefont {H.}~\bibnamefont {Wu}},\ }\bibfield  {title} {\bibinfo
  {title} {Physical reservoir computing with emerging electronics},\ }\href
  {https://doi.org/10.1038/s41928-024-01133-z} {\bibfield  {journal} {\bibinfo
  {journal} {Nat. Electron.}\ }\textbf {\bibinfo {volume} {7}},\ \bibinfo
  {pages} {193} (\bibinfo {year} {2024})}\BibitemShut {NoStop}%
\bibitem [{\citenamefont {Jaeger}\ and\ \citenamefont
  {Haas}(2004)}]{jaegerHarnessingNonlinearityPredicting2004}%
  \BibitemOpen
  \bibfield  {author} {\bibinfo {author} {\bibfnamefont {H.}~\bibnamefont
  {Jaeger}}\ and\ \bibinfo {author} {\bibfnamefont {H.}~\bibnamefont {Haas}},\
  }\bibfield  {title} {\bibinfo {title} {Harnessing {{Nonlinearity}}:
  {{Predicting Chaotic Systems}} and {{Saving Energy}} in {{Wireless
  Communication}}},\ }\href {https://doi.org/10.1126/science.1091277}
  {\bibfield  {journal} {\bibinfo  {journal} {Science}\ }\textbf {\bibinfo
  {volume} {304}},\ \bibinfo {pages} {78} (\bibinfo {year} {2004})}\BibitemShut
  {NoStop}%
\bibitem [{\citenamefont {Maass}\ \emph {et~al.}(2002)\citenamefont {Maass},
  \citenamefont {Natschl{\"a}ger},\ and\ \citenamefont
  {Markram}}]{10.1162/089976602760407955}%
  \BibitemOpen
  \bibfield  {author} {\bibinfo {author} {\bibfnamefont {W.}~\bibnamefont
  {Maass}}, \bibinfo {author} {\bibfnamefont {T.}~\bibnamefont
  {Natschl{\"a}ger}},\ and\ \bibinfo {author} {\bibfnamefont {H.}~\bibnamefont
  {Markram}},\ }\bibfield  {title} {\bibinfo {title} {Real-time computing
  without stable states: A new framework for neural computation based on
  perturbations},\ }\href {https://doi.org/10.1162/089976602760407955}
  {\bibfield  {journal} {\bibinfo  {journal} {Neural Computation}\ }\textbf
  {\bibinfo {volume} {14}},\ \bibinfo {pages} {2531} (\bibinfo {year}
  {2002})},\ \Eprint
  {https://arxiv.org/abs/https://doi.org/10.1162/089976602760407955}
  {https://doi.org/10.1162/089976602760407955} \BibitemShut {NoStop}%
\bibitem [{\citenamefont {Verstraeten}\ \emph {et~al.}(2007)\citenamefont
  {Verstraeten}, \citenamefont {Schrauwen}, \citenamefont {D'Haene},\ and\
  \citenamefont
  {Stroobandt}}]{verstraetenExperimentalUnificationReservoir2007}%
  \BibitemOpen
  \bibfield  {author} {\bibinfo {author} {\bibfnamefont {D.}~\bibnamefont
  {Verstraeten}}, \bibinfo {author} {\bibfnamefont {B.}~\bibnamefont
  {Schrauwen}}, \bibinfo {author} {\bibfnamefont {M.}~\bibnamefont {D'Haene}},\
  and\ \bibinfo {author} {\bibfnamefont {D.}~\bibnamefont {Stroobandt}},\
  }\bibfield  {title} {\bibinfo {title} {An experimental unification of
  reservoir computing methods},\ }\href
  {https://doi.org/10.1016/j.neunet.2007.04.003} {\bibfield  {journal}
  {\bibinfo  {journal} {Neural Networks}\ }\textbf {\bibinfo {volume} {20}},\
  \bibinfo {pages} {391} (\bibinfo {year} {2007})}\BibitemShut {NoStop}%
\bibitem [{\citenamefont {Nakajima}\ and\ \citenamefont
  {Fischer}(2021)}]{nakajimaReservoirComputingTheory2021}%
  \BibitemOpen
  \bibinfo {editor} {\bibfnamefont {K.}~\bibnamefont {Nakajima}}\ and\ \bibinfo
  {editor} {\bibfnamefont {I.}~\bibnamefont {Fischer}},\ eds.,\ \href
  {https://doi.org/10.1007/978-981-13-1687-6} {\emph {\bibinfo {title}
  {Reservoir {{Computing}}: {{Theory}}, {{Physical Implementations}}, and
  {{Applications}}}}},\ Natural {{Computing Series}}\ (\bibinfo  {publisher}
  {Springer Singapore},\ \bibinfo {address} {Singapore},\ \bibinfo {year}
  {2021})\BibitemShut {NoStop}%
\bibitem [{\citenamefont {Duport}\ \emph {et~al.}(2012)\citenamefont {Duport},
  \citenamefont {Schneider}, \citenamefont {Smerieri}, \citenamefont
  {Haelterman},\ and\ \citenamefont
  {Massar}}]{duportAllopticalReservoirComputing2012}%
  \BibitemOpen
  \bibfield  {author} {\bibinfo {author} {\bibfnamefont {F.}~\bibnamefont
  {Duport}}, \bibinfo {author} {\bibfnamefont {B.}~\bibnamefont {Schneider}},
  \bibinfo {author} {\bibfnamefont {A.}~\bibnamefont {Smerieri}}, \bibinfo
  {author} {\bibfnamefont {M.}~\bibnamefont {Haelterman}},\ and\ \bibinfo
  {author} {\bibfnamefont {S.}~\bibnamefont {Massar}},\ }\bibfield  {title}
  {\bibinfo {title} {All-optical reservoir computing},\ }\href
  {https://doi.org/10.1364/OE.20.022783} {\bibfield  {journal} {\bibinfo
  {journal} {Optics Express}\ }\textbf {\bibinfo {volume} {20}},\ \bibinfo
  {pages} {22783} (\bibinfo {year} {2012})}\BibitemShut {NoStop}%
\bibitem [{\citenamefont {Van Der~Sande}\ \emph {et~al.}(2017)\citenamefont
  {Van Der~Sande}, \citenamefont {Brunner},\ and\ \citenamefont
  {Soriano}}]{vandersandeAdvancesPhotonicReservoir2017}%
  \BibitemOpen
  \bibfield  {author} {\bibinfo {author} {\bibfnamefont {G.}~\bibnamefont {Van
  Der~Sande}}, \bibinfo {author} {\bibfnamefont {D.}~\bibnamefont {Brunner}},\
  and\ \bibinfo {author} {\bibfnamefont {M.~C.}\ \bibnamefont {Soriano}},\
  }\bibfield  {title} {\bibinfo {title} {Advances in photonic reservoir
  computing},\ }\href {https://doi.org/10.1515/nanoph-2016-0132} {\bibfield
  {journal} {\bibinfo  {journal} {Nanophotonics}\ }\textbf {\bibinfo {volume}
  {6}},\ \bibinfo {pages} {561} (\bibinfo {year} {2017})}\BibitemShut {NoStop}%
\bibitem [{\citenamefont {Chembo}(2020)}]{chemboMachineLearningBased2020}%
  \BibitemOpen
  \bibfield  {author} {\bibinfo {author} {\bibfnamefont {Y.~K.}\ \bibnamefont
  {Chembo}},\ }\bibfield  {title} {\bibinfo {title} {Machine learning based on
  reservoir computing with time-delayed optoelectronic and photonic systems},\
  }\href {https://doi.org/10.1063/1.5120788} {\bibfield  {journal} {\bibinfo
  {journal} {Chaos: An Interdisciplinary Journal of Nonlinear Science}\
  }\textbf {\bibinfo {volume} {30}},\ \bibinfo {pages} {013111} (\bibinfo
  {year} {2020})}\BibitemShut {NoStop}%
\bibitem [{\citenamefont {Yan}\ \emph {et~al.}(2024)\citenamefont {Yan},
  \citenamefont {Huang}, \citenamefont {Bienstman}, \citenamefont {Tino},
  \citenamefont {Lin},\ and\ \citenamefont
  {Sun}}]{yanEmergingOpportunitiesChallenges2024}%
  \BibitemOpen
  \bibfield  {author} {\bibinfo {author} {\bibfnamefont {M.}~\bibnamefont
  {Yan}}, \bibinfo {author} {\bibfnamefont {C.}~\bibnamefont {Huang}}, \bibinfo
  {author} {\bibfnamefont {P.}~\bibnamefont {Bienstman}}, \bibinfo {author}
  {\bibfnamefont {P.}~\bibnamefont {Tino}}, \bibinfo {author} {\bibfnamefont
  {W.}~\bibnamefont {Lin}},\ and\ \bibinfo {author} {\bibfnamefont
  {J.}~\bibnamefont {Sun}},\ }\bibfield  {title} {\bibinfo {title} {Emerging
  opportunities and challenges for the future of reservoir computing},\ }\href
  {https://doi.org/10.1038/s41467-024-45187-1} {\bibfield  {journal} {\bibinfo
  {journal} {Nat. Commun.}\ }\textbf {\bibinfo {volume} {15}},\ \bibinfo
  {pages} {2056} (\bibinfo {year} {2024})}\BibitemShut {NoStop}%
\bibitem [{\citenamefont {Du}\ \emph {et~al.}(2017)\citenamefont {Du},
  \citenamefont {Cai}, \citenamefont {Zidan}, \citenamefont {Ma}, \citenamefont
  {Lee},\ and\ \citenamefont {Lu}}]{duReservoirComputingUsing2017}%
  \BibitemOpen
  \bibfield  {author} {\bibinfo {author} {\bibfnamefont {C.}~\bibnamefont
  {Du}}, \bibinfo {author} {\bibfnamefont {F.}~\bibnamefont {Cai}}, \bibinfo
  {author} {\bibfnamefont {M.~A.}\ \bibnamefont {Zidan}}, \bibinfo {author}
  {\bibfnamefont {W.}~\bibnamefont {Ma}}, \bibinfo {author} {\bibfnamefont
  {S.~H.}\ \bibnamefont {Lee}},\ and\ \bibinfo {author} {\bibfnamefont {W.~D.}\
  \bibnamefont {Lu}},\ }\bibfield  {title} {\bibinfo {title} {Reservoir
  computing using dynamic memristors for temporal information processing},\
  }\href {https://doi.org/10.1038/s41467-017-02337-y} {\bibfield  {journal}
  {\bibinfo  {journal} {Nat. Commun.}\ }\textbf {\bibinfo {volume} {8}},\
  \bibinfo {pages} {2204} (\bibinfo {year} {2017})}\BibitemShut {NoStop}%
\bibitem [{\citenamefont {Moon}\ \emph {et~al.}(2019)\citenamefont {Moon},
  \citenamefont {Ma}, \citenamefont {Shin}, \citenamefont {Cai}, \citenamefont
  {Du}, \citenamefont {Lee},\ and\ \citenamefont
  {Lu}}]{moonTemporalDataClassification2019}%
  \BibitemOpen
  \bibfield  {author} {\bibinfo {author} {\bibfnamefont {J.}~\bibnamefont
  {Moon}}, \bibinfo {author} {\bibfnamefont {W.}~\bibnamefont {Ma}}, \bibinfo
  {author} {\bibfnamefont {J.~H.}\ \bibnamefont {Shin}}, \bibinfo {author}
  {\bibfnamefont {F.}~\bibnamefont {Cai}}, \bibinfo {author} {\bibfnamefont
  {C.}~\bibnamefont {Du}}, \bibinfo {author} {\bibfnamefont {S.~H.}\
  \bibnamefont {Lee}},\ and\ \bibinfo {author} {\bibfnamefont {W.~D.}\
  \bibnamefont {Lu}},\ }\bibfield  {title} {\bibinfo {title} {Temporal data
  classification and forecasting using a memristor-based reservoir computing
  system},\ }\href {https://doi.org/10.1038/s41928-019-0313-3} {\bibfield
  {journal} {\bibinfo  {journal} {Nat. Electron.}\ }\textbf {\bibinfo {volume}
  {2}},\ \bibinfo {pages} {480} (\bibinfo {year} {2019})}\BibitemShut {NoStop}%
\bibitem [{\citenamefont {Zhong}\ \emph {et~al.}(2021)\citenamefont {Zhong},
  \citenamefont {Tang}, \citenamefont {Li}, \citenamefont {Gao}, \citenamefont
  {Qian},\ and\ \citenamefont {Wu}}]{zhongDynamicMemristorbasedReservoir2021}%
  \BibitemOpen
  \bibfield  {author} {\bibinfo {author} {\bibfnamefont {Y.}~\bibnamefont
  {Zhong}}, \bibinfo {author} {\bibfnamefont {J.}~\bibnamefont {Tang}},
  \bibinfo {author} {\bibfnamefont {X.}~\bibnamefont {Li}}, \bibinfo {author}
  {\bibfnamefont {B.}~\bibnamefont {Gao}}, \bibinfo {author} {\bibfnamefont
  {H.}~\bibnamefont {Qian}},\ and\ \bibinfo {author} {\bibfnamefont
  {H.}~\bibnamefont {Wu}},\ }\bibfield  {title} {\bibinfo {title} {Dynamic
  memristor-based reservoir computing for high-efficiency temporal signal
  processing},\ }\href {https://doi.org/10.1038/s41467-020-20692-1} {\bibfield
  {journal} {\bibinfo  {journal} {Nat. Commun.}\ }\textbf {\bibinfo {volume}
  {12}},\ \bibinfo {pages} {408} (\bibinfo {year} {2021})}\BibitemShut
  {NoStop}%
\bibitem [{\citenamefont {Zhong}\ \emph {et~al.}(2022)\citenamefont {Zhong},
  \citenamefont {Tang}, \citenamefont {Li}, \citenamefont {Liang},
  \citenamefont {Liu}, \citenamefont {Li}, \citenamefont {Xi}, \citenamefont
  {Yao}, \citenamefont {Hao}, \citenamefont {Gao}, \citenamefont {Qian},\ and\
  \citenamefont {Wu}}]{zhongMemristorbasedAnalogueReservoir2022}%
  \BibitemOpen
  \bibfield  {author} {\bibinfo {author} {\bibfnamefont {Y.}~\bibnamefont
  {Zhong}}, \bibinfo {author} {\bibfnamefont {J.}~\bibnamefont {Tang}},
  \bibinfo {author} {\bibfnamefont {X.}~\bibnamefont {Li}}, \bibinfo {author}
  {\bibfnamefont {X.}~\bibnamefont {Liang}}, \bibinfo {author} {\bibfnamefont
  {Z.}~\bibnamefont {Liu}}, \bibinfo {author} {\bibfnamefont {Y.}~\bibnamefont
  {Li}}, \bibinfo {author} {\bibfnamefont {Y.}~\bibnamefont {Xi}}, \bibinfo
  {author} {\bibfnamefont {P.}~\bibnamefont {Yao}}, \bibinfo {author}
  {\bibfnamefont {Z.}~\bibnamefont {Hao}}, \bibinfo {author} {\bibfnamefont
  {B.}~\bibnamefont {Gao}}, \bibinfo {author} {\bibfnamefont {H.}~\bibnamefont
  {Qian}},\ and\ \bibinfo {author} {\bibfnamefont {H.}~\bibnamefont {Wu}},\
  }\bibfield  {title} {\bibinfo {title} {A memristor-based analogue reservoir
  computing system for real-time and power-efficient signal processing},\
  }\href {https://doi.org/10.1038/s41928-022-00838-3} {\bibfield  {journal}
  {\bibinfo  {journal} {Nat. Electron.}\ }\textbf {\bibinfo {volume} {5}},\
  \bibinfo {pages} {672} (\bibinfo {year} {2022})}\BibitemShut {NoStop}%
\bibitem [{\citenamefont {Kanao}\ \emph {et~al.}(2019)\citenamefont {Kanao},
  \citenamefont {Suto}, \citenamefont {Mizushima}, \citenamefont {Goto},
  \citenamefont {Tanamoto},\ and\ \citenamefont
  {Nagasawa}}]{kanaoReservoirComputingSpinTorque2019}%
  \BibitemOpen
  \bibfield  {author} {\bibinfo {author} {\bibfnamefont {T.}~\bibnamefont
  {Kanao}}, \bibinfo {author} {\bibfnamefont {H.}~\bibnamefont {Suto}},
  \bibinfo {author} {\bibfnamefont {K.}~\bibnamefont {Mizushima}}, \bibinfo
  {author} {\bibfnamefont {H.}~\bibnamefont {Goto}}, \bibinfo {author}
  {\bibfnamefont {T.}~\bibnamefont {Tanamoto}},\ and\ \bibinfo {author}
  {\bibfnamefont {T.}~\bibnamefont {Nagasawa}},\ }\bibfield  {title} {\bibinfo
  {title} {Reservoir {{Computing}} on {{Spin-Torque Oscillator Array}}},\
  }\href {https://doi.org/10.1103/PhysRevApplied.12.024052} {\bibfield
  {journal} {\bibinfo  {journal} {Phys. Rev. Appl.}\ }\textbf {\bibinfo
  {volume} {12}},\ \bibinfo {pages} {024052} (\bibinfo {year}
  {2019})}\BibitemShut {NoStop}%
\bibitem [{\citenamefont {Tsunegi}\ \emph {et~al.}(2019)\citenamefont
  {Tsunegi}, \citenamefont {Taniguchi}, \citenamefont {Nakajima}, \citenamefont
  {Miwa}, \citenamefont {Yakushiji}, \citenamefont {Fukushima}, \citenamefont
  {Yuasa},\ and\ \citenamefont
  {Kubota}}]{tsunegiPhysicalReservoirComputing2019}%
  \BibitemOpen
  \bibfield  {author} {\bibinfo {author} {\bibfnamefont {S.}~\bibnamefont
  {Tsunegi}}, \bibinfo {author} {\bibfnamefont {T.}~\bibnamefont {Taniguchi}},
  \bibinfo {author} {\bibfnamefont {K.}~\bibnamefont {Nakajima}}, \bibinfo
  {author} {\bibfnamefont {S.}~\bibnamefont {Miwa}}, \bibinfo {author}
  {\bibfnamefont {K.}~\bibnamefont {Yakushiji}}, \bibinfo {author}
  {\bibfnamefont {A.}~\bibnamefont {Fukushima}}, \bibinfo {author}
  {\bibfnamefont {S.}~\bibnamefont {Yuasa}},\ and\ \bibinfo {author}
  {\bibfnamefont {H.}~\bibnamefont {Kubota}},\ }\bibfield  {title} {\bibinfo
  {title} {Physical reservoir computing based on spin torque oscillator with
  forced synchronization},\ }\href {https://doi.org/10.1063/1.5081797}
  {\bibfield  {journal} {\bibinfo  {journal} {Appl. Phys. Lett.}\ }\textbf
  {\bibinfo {volume} {114}},\ \bibinfo {pages} {164101} (\bibinfo {year}
  {2019})}\BibitemShut {NoStop}%
\bibitem [{\citenamefont {Markovi{\'c}}\ \emph {et~al.}(2019)\citenamefont
  {Markovi{\'c}}, \citenamefont {Leroux}, \citenamefont {Riou}, \citenamefont
  {Abreu~Araujo}, \citenamefont {Torrejon}, \citenamefont {Querlioz},
  \citenamefont {Fukushima}, \citenamefont {Yuasa}, \citenamefont {Trastoy},
  \citenamefont {Bortolotti},\ and\ \citenamefont
  {Grollier}}]{markovicReservoirComputingFrequency2019}%
  \BibitemOpen
  \bibfield  {author} {\bibinfo {author} {\bibfnamefont {D.}~\bibnamefont
  {Markovi{\'c}}}, \bibinfo {author} {\bibfnamefont {N.}~\bibnamefont
  {Leroux}}, \bibinfo {author} {\bibfnamefont {M.}~\bibnamefont {Riou}},
  \bibinfo {author} {\bibfnamefont {F.}~\bibnamefont {Abreu~Araujo}}, \bibinfo
  {author} {\bibfnamefont {J.}~\bibnamefont {Torrejon}}, \bibinfo {author}
  {\bibfnamefont {D.}~\bibnamefont {Querlioz}}, \bibinfo {author}
  {\bibfnamefont {A.}~\bibnamefont {Fukushima}}, \bibinfo {author}
  {\bibfnamefont {S.}~\bibnamefont {Yuasa}}, \bibinfo {author} {\bibfnamefont
  {J.}~\bibnamefont {Trastoy}}, \bibinfo {author} {\bibfnamefont
  {P.}~\bibnamefont {Bortolotti}},\ and\ \bibinfo {author} {\bibfnamefont
  {J.}~\bibnamefont {Grollier}},\ }\bibfield  {title} {\bibinfo {title}
  {Reservoir computing with the frequency, phase, and amplitude of spin-torque
  nano-oscillators},\ }\href {https://doi.org/10.1063/1.5079305} {\bibfield
  {journal} {\bibinfo  {journal} {Appl. Phys. Lett.}\ }\textbf {\bibinfo
  {volume} {114}},\ \bibinfo {pages} {012409} (\bibinfo {year}
  {2019})}\BibitemShut {NoStop}%
\bibitem [{\citenamefont {Yamaguchi}\ \emph {et~al.}(2020)\citenamefont
  {Yamaguchi}, \citenamefont {Akashi}, \citenamefont {Nakajima}, \citenamefont
  {Kubota}, \citenamefont {Tsunegi},\ and\ \citenamefont
  {Taniguchi}}]{yamaguchi2020step}%
  \BibitemOpen
  \bibfield  {author} {\bibinfo {author} {\bibfnamefont {T.}~\bibnamefont
  {Yamaguchi}}, \bibinfo {author} {\bibfnamefont {N.}~\bibnamefont {Akashi}},
  \bibinfo {author} {\bibfnamefont {K.}~\bibnamefont {Nakajima}}, \bibinfo
  {author} {\bibfnamefont {H.}~\bibnamefont {Kubota}}, \bibinfo {author}
  {\bibfnamefont {S.}~\bibnamefont {Tsunegi}},\ and\ \bibinfo {author}
  {\bibfnamefont {T.}~\bibnamefont {Taniguchi}},\ }\bibfield  {title} {\bibinfo
  {title} {Step-like dependence of memory function on pulse width in
  spintronics reservoir computing},\ }\href
  {https://doi.org/10.1038/s41598-020-76142-x} {\bibfield  {journal} {\bibinfo
  {journal} {Sci. Rep.}\ }\textbf {\bibinfo {volume} {10}},\ \bibinfo {pages}
  {19536} (\bibinfo {year} {2020})}\BibitemShut {NoStop}%
\bibitem [{\citenamefont {Taniguchi}\ \emph {et~al.}(2022)\citenamefont
  {Taniguchi}, \citenamefont {Ogihara}, \citenamefont {Utsumi},\ and\
  \citenamefont {Tsunegi}}]{taniguchiSpintronicReservoirComputing2022}%
  \BibitemOpen
  \bibfield  {author} {\bibinfo {author} {\bibfnamefont {T.}~\bibnamefont
  {Taniguchi}}, \bibinfo {author} {\bibfnamefont {A.}~\bibnamefont {Ogihara}},
  \bibinfo {author} {\bibfnamefont {Y.}~\bibnamefont {Utsumi}},\ and\ \bibinfo
  {author} {\bibfnamefont {S.}~\bibnamefont {Tsunegi}},\ }\bibfield  {title}
  {\bibinfo {title} {Spintronic reservoir computing without driving current or
  magnetic field},\ }\href {https://doi.org/10.1038/s41598-022-14738-1}
  {\bibfield  {journal} {\bibinfo  {journal} {Sci. Rep.}\ }\textbf {\bibinfo
  {volume} {12}},\ \bibinfo {pages} {10627} (\bibinfo {year}
  {2022})}\BibitemShut {NoStop}%
\bibitem [{\citenamefont {Akashi}\ \emph {et~al.}(2022)\citenamefont {Akashi},
  \citenamefont {Kuniyoshi}, \citenamefont {Tsunegi}, \citenamefont
  {Taniguchi}, \citenamefont {Nishida}, \citenamefont {Sakurai}, \citenamefont
  {Wakao}, \citenamefont {Kawashima},\ and\ \citenamefont
  {Nakajima}}]{akashiCoupledSpintronicsNeuromorphic2022}%
  \BibitemOpen
  \bibfield  {author} {\bibinfo {author} {\bibfnamefont {N.}~\bibnamefont
  {Akashi}}, \bibinfo {author} {\bibfnamefont {Y.}~\bibnamefont {Kuniyoshi}},
  \bibinfo {author} {\bibfnamefont {S.}~\bibnamefont {Tsunegi}}, \bibinfo
  {author} {\bibfnamefont {T.}~\bibnamefont {Taniguchi}}, \bibinfo {author}
  {\bibfnamefont {M.}~\bibnamefont {Nishida}}, \bibinfo {author} {\bibfnamefont
  {R.}~\bibnamefont {Sakurai}}, \bibinfo {author} {\bibfnamefont
  {Y.}~\bibnamefont {Wakao}}, \bibinfo {author} {\bibfnamefont
  {K.}~\bibnamefont {Kawashima}},\ and\ \bibinfo {author} {\bibfnamefont
  {K.}~\bibnamefont {Nakajima}},\ }\bibfield  {title} {\bibinfo {title} {A
  {{Coupled Spintronics Neuromorphic Approach}} for {{High}}-{{Performance
  Reservoir Computing}}},\ }\href {https://doi.org/10.1002/aisy.202200123}
  {\bibfield  {journal} {\bibinfo  {journal} {Advanced Intelligent Systems}\
  }\textbf {\bibinfo {volume} {4}},\ \bibinfo {pages} {2200123} (\bibinfo
  {year} {2022})}\BibitemShut {NoStop}%
\bibitem [{\citenamefont {Gartside}\ \emph {et~al.}(2022)\citenamefont
  {Gartside}, \citenamefont {Stenning}, \citenamefont {Vanstone}, \citenamefont
  {Holder}, \citenamefont {Arroo}, \citenamefont {Dion}, \citenamefont
  {Caravelli}, \citenamefont {Kurebayashi},\ and\ \citenamefont
  {Branford}}]{gartsideReconfigurableTrainingReservoir2022}%
  \BibitemOpen
  \bibfield  {author} {\bibinfo {author} {\bibfnamefont {J.~C.}\ \bibnamefont
  {Gartside}}, \bibinfo {author} {\bibfnamefont {K.~D.}\ \bibnamefont
  {Stenning}}, \bibinfo {author} {\bibfnamefont {A.}~\bibnamefont {Vanstone}},
  \bibinfo {author} {\bibfnamefont {H.~H.}\ \bibnamefont {Holder}}, \bibinfo
  {author} {\bibfnamefont {D.~M.}\ \bibnamefont {Arroo}}, \bibinfo {author}
  {\bibfnamefont {T.}~\bibnamefont {Dion}}, \bibinfo {author} {\bibfnamefont
  {F.}~\bibnamefont {Caravelli}}, \bibinfo {author} {\bibfnamefont
  {H.}~\bibnamefont {Kurebayashi}},\ and\ \bibinfo {author} {\bibfnamefont
  {W.~R.}\ \bibnamefont {Branford}},\ }\bibfield  {title} {\bibinfo {title}
  {Reconfigurable training and reservoir computing in an artificial spin-vortex
  ice via spin-wave fingerprinting},\ }\href
  {https://doi.org/10.1038/s41565-022-01091-7} {\bibfield  {journal} {\bibinfo
  {journal} {Nat. Nanotechnol.}\ }\textbf {\bibinfo {volume} {17}},\ \bibinfo
  {pages} {460} (\bibinfo {year} {2022})}\BibitemShut {NoStop}%
\bibitem [{\citenamefont {Allwood}\ \emph {et~al.}(2023)\citenamefont
  {Allwood}, \citenamefont {Ellis}, \citenamefont {Griffin}, \citenamefont
  {Hayward}, \citenamefont {Manneschi}, \citenamefont {Musameh}, \citenamefont
  {O'Keefe}, \citenamefont {Stepney}, \citenamefont {Swindells}, \citenamefont
  {Trefzer}, \citenamefont {Vasilaki}, \citenamefont {Venkat}, \citenamefont
  {Vidamour},\ and\ \citenamefont
  {Wringe}}]{allwoodPerspectivePhysicalReservoir2023}%
  \BibitemOpen
  \bibfield  {author} {\bibinfo {author} {\bibfnamefont {D.~A.}\ \bibnamefont
  {Allwood}}, \bibinfo {author} {\bibfnamefont {M.~O.~A.}\ \bibnamefont
  {Ellis}}, \bibinfo {author} {\bibfnamefont {D.}~\bibnamefont {Griffin}},
  \bibinfo {author} {\bibfnamefont {T.~J.}\ \bibnamefont {Hayward}}, \bibinfo
  {author} {\bibfnamefont {L.}~\bibnamefont {Manneschi}}, \bibinfo {author}
  {\bibfnamefont {M.~F.~K.}\ \bibnamefont {Musameh}}, \bibinfo {author}
  {\bibfnamefont {S.}~\bibnamefont {O'Keefe}}, \bibinfo {author} {\bibfnamefont
  {S.}~\bibnamefont {Stepney}}, \bibinfo {author} {\bibfnamefont
  {C.}~\bibnamefont {Swindells}}, \bibinfo {author} {\bibfnamefont {M.~A.}\
  \bibnamefont {Trefzer}}, \bibinfo {author} {\bibfnamefont {E.}~\bibnamefont
  {Vasilaki}}, \bibinfo {author} {\bibfnamefont {G.}~\bibnamefont {Venkat}},
  \bibinfo {author} {\bibfnamefont {I.}~\bibnamefont {Vidamour}},\ and\
  \bibinfo {author} {\bibfnamefont {C.}~\bibnamefont {Wringe}},\ }\bibfield
  {title} {\bibinfo {title} {A perspective on physical reservoir computing with
  nanomagnetic devices},\ }\href {https://doi.org/10.1063/5.0119040} {\bibfield
   {journal} {\bibinfo  {journal} {Appl. Phys. Lett.}\ }\textbf {\bibinfo
  {volume} {122}},\ \bibinfo {pages} {040501} (\bibinfo {year}
  {2023})}\BibitemShut {NoStop}%
\bibitem [{\citenamefont {Jensen}\ \emph {et~al.}(2024)\citenamefont {Jensen},
  \citenamefont {Str{\o}mberg}, \citenamefont {Breivik}, \citenamefont {Penty},
  \citenamefont {Ni{\~n}o}, \citenamefont {Khaliq}, \citenamefont {Foerster},
  \citenamefont {Tufte},\ and\ \citenamefont
  {Folven}}]{jensenClockedDynamicsArtificial2024}%
  \BibitemOpen
  \bibfield  {author} {\bibinfo {author} {\bibfnamefont {J.~H.}\ \bibnamefont
  {Jensen}}, \bibinfo {author} {\bibfnamefont {A.}~\bibnamefont
  {Str{\o}mberg}}, \bibinfo {author} {\bibfnamefont {I.}~\bibnamefont
  {Breivik}}, \bibinfo {author} {\bibfnamefont {A.}~\bibnamefont {Penty}},
  \bibinfo {author} {\bibfnamefont {M.~A.}\ \bibnamefont {Ni{\~n}o}}, \bibinfo
  {author} {\bibfnamefont {M.~W.}\ \bibnamefont {Khaliq}}, \bibinfo {author}
  {\bibfnamefont {M.}~\bibnamefont {Foerster}}, \bibinfo {author}
  {\bibfnamefont {G.}~\bibnamefont {Tufte}},\ and\ \bibinfo {author}
  {\bibfnamefont {E.}~\bibnamefont {Folven}},\ }\bibfield  {title} {\bibinfo
  {title} {Clocked dynamics in artificial spin ice},\ }\href
  {https://doi.org/10.1038/s41467-024-45319-7} {\bibfield  {journal} {\bibinfo
  {journal} {Nat. Commun.}\ }\textbf {\bibinfo {volume} {15}},\ \bibinfo
  {pages} {964} (\bibinfo {year} {2024})}\BibitemShut {NoStop}%
\bibitem [{\citenamefont {Prychynenko}\ \emph {et~al.}(2018)\citenamefont
  {Prychynenko}, \citenamefont {Sitte}, \citenamefont {Litzius}, \citenamefont
  {Kr{\"u}ger}, \citenamefont {Bourianoff}, \citenamefont {Kl{\"a}ui},
  \citenamefont {Sinova},\ and\ \citenamefont
  {{Everschor-Sitte}}}]{prychynenkoMagneticSkyrmionNonlinear2018}%
  \BibitemOpen
  \bibfield  {author} {\bibinfo {author} {\bibfnamefont {D.}~\bibnamefont
  {Prychynenko}}, \bibinfo {author} {\bibfnamefont {M.}~\bibnamefont {Sitte}},
  \bibinfo {author} {\bibfnamefont {K.}~\bibnamefont {Litzius}}, \bibinfo
  {author} {\bibfnamefont {B.}~\bibnamefont {Kr{\"u}ger}}, \bibinfo {author}
  {\bibfnamefont {G.}~\bibnamefont {Bourianoff}}, \bibinfo {author}
  {\bibfnamefont {M.}~\bibnamefont {Kl{\"a}ui}}, \bibinfo {author}
  {\bibfnamefont {J.}~\bibnamefont {Sinova}},\ and\ \bibinfo {author}
  {\bibfnamefont {K.}~\bibnamefont {{Everschor-Sitte}}},\ }\bibfield  {title}
  {\bibinfo {title} {Magnetic {{Skyrmion}} as a {{Nonlinear Resistive
  Element}}: {{A Potential Building Block}} for {{Reservoir Computing}}},\
  }\href {https://doi.org/10.1103/PhysRevApplied.9.014034} {\bibfield
  {journal} {\bibinfo  {journal} {Phys. Rev. Appl.}\ }\textbf {\bibinfo
  {volume} {9}},\ \bibinfo {pages} {014034} (\bibinfo {year}
  {2018})}\BibitemShut {NoStop}%
\bibitem [{\citenamefont {Pinna}\ \emph {et~al.}(2020)\citenamefont {Pinna},
  \citenamefont {Bourianoff},\ and\ \citenamefont
  {{Everschor-Sitte}}}]{pinnaReservoirComputingRandom2020}%
  \BibitemOpen
  \bibfield  {author} {\bibinfo {author} {\bibfnamefont {D.}~\bibnamefont
  {Pinna}}, \bibinfo {author} {\bibfnamefont {G.}~\bibnamefont {Bourianoff}},\
  and\ \bibinfo {author} {\bibfnamefont {K.}~\bibnamefont
  {{Everschor-Sitte}}},\ }\bibfield  {title} {\bibinfo {title} {Reservoir
  {{Computing}} with {{Random Skyrmion Textures}}},\ }\href
  {https://doi.org/10.1103/PhysRevApplied.14.054020} {\bibfield  {journal}
  {\bibinfo  {journal} {Phys. Rev. Appl.}\ }\textbf {\bibinfo {volume} {14}},\
  \bibinfo {pages} {054020} (\bibinfo {year} {2020})}\BibitemShut {NoStop}%
\bibitem [{\citenamefont {Sun}\ \emph {et~al.}(2023)\citenamefont {Sun},
  \citenamefont {Lin}, \citenamefont {Lei}, \citenamefont {Chen}, \citenamefont
  {Kang}, \citenamefont {Zhao}, \citenamefont {Wei}, \citenamefont {Chen},
  \citenamefont {Pang}, \citenamefont {Hu}, \citenamefont {Yang}, \citenamefont
  {Dong}, \citenamefont {Zhao}, \citenamefont {Liu}, \citenamefont {Yuan},
  \citenamefont {Ullrich}, \citenamefont {Back}, \citenamefont {Zhang},
  \citenamefont {Pan}, \citenamefont {Zhao}, \citenamefont {Feng},
  \citenamefont {Fert},\ and\ \citenamefont
  {Zhao}}]{sunExperimentalDemonstrationSkyrmionenhanced2023}%
  \BibitemOpen
  \bibfield  {author} {\bibinfo {author} {\bibfnamefont {Y.}~\bibnamefont
  {Sun}}, \bibinfo {author} {\bibfnamefont {T.}~\bibnamefont {Lin}}, \bibinfo
  {author} {\bibfnamefont {N.}~\bibnamefont {Lei}}, \bibinfo {author}
  {\bibfnamefont {X.}~\bibnamefont {Chen}}, \bibinfo {author} {\bibfnamefont
  {W.}~\bibnamefont {Kang}}, \bibinfo {author} {\bibfnamefont {Z.}~\bibnamefont
  {Zhao}}, \bibinfo {author} {\bibfnamefont {D.}~\bibnamefont {Wei}}, \bibinfo
  {author} {\bibfnamefont {C.}~\bibnamefont {Chen}}, \bibinfo {author}
  {\bibfnamefont {S.}~\bibnamefont {Pang}}, \bibinfo {author} {\bibfnamefont
  {L.}~\bibnamefont {Hu}}, \bibinfo {author} {\bibfnamefont {L.}~\bibnamefont
  {Yang}}, \bibinfo {author} {\bibfnamefont {E.}~\bibnamefont {Dong}}, \bibinfo
  {author} {\bibfnamefont {L.}~\bibnamefont {Zhao}}, \bibinfo {author}
  {\bibfnamefont {L.}~\bibnamefont {Liu}}, \bibinfo {author} {\bibfnamefont
  {Z.}~\bibnamefont {Yuan}}, \bibinfo {author} {\bibfnamefont {A.}~\bibnamefont
  {Ullrich}}, \bibinfo {author} {\bibfnamefont {C.~H.}\ \bibnamefont {Back}},
  \bibinfo {author} {\bibfnamefont {J.}~\bibnamefont {Zhang}}, \bibinfo
  {author} {\bibfnamefont {D.}~\bibnamefont {Pan}}, \bibinfo {author}
  {\bibfnamefont {J.}~\bibnamefont {Zhao}}, \bibinfo {author} {\bibfnamefont
  {M.}~\bibnamefont {Feng}}, \bibinfo {author} {\bibfnamefont {A.}~\bibnamefont
  {Fert}},\ and\ \bibinfo {author} {\bibfnamefont {W.}~\bibnamefont {Zhao}},\
  }\bibfield  {title} {\bibinfo {title} {Experimental demonstration of a
  skyrmion-enhanced strain-mediated physical reservoir computing system},\
  }\href {https://doi.org/10.1038/s41467-023-39207-9} {\bibfield  {journal}
  {\bibinfo  {journal} {Nat. Commun.}\ }\textbf {\bibinfo {volume} {14}},\
  \bibinfo {pages} {3434} (\bibinfo {year} {2023})}\BibitemShut {NoStop}%
\bibitem [{\citenamefont {Nakane}\ \emph {et~al.}(2018)\citenamefont {Nakane},
  \citenamefont {Tanaka},\ and\ \citenamefont
  {Hirose}}]{nakaneReservoirComputingSpin2018}%
  \BibitemOpen
  \bibfield  {author} {\bibinfo {author} {\bibfnamefont {R.}~\bibnamefont
  {Nakane}}, \bibinfo {author} {\bibfnamefont {G.}~\bibnamefont {Tanaka}},\
  and\ \bibinfo {author} {\bibfnamefont {A.}~\bibnamefont {Hirose}},\
  }\bibfield  {title} {\bibinfo {title} {Reservoir {{Computing With Spin Waves
  Excited}} in a {{Garnet Film}}},\ }\href
  {https://doi.org/10.1109/ACCESS.2018.2794584} {\bibfield  {journal} {\bibinfo
   {journal} {IEEE Access}\ }\textbf {\bibinfo {volume} {6}},\ \bibinfo {pages}
  {4462} (\bibinfo {year} {2018})}\BibitemShut {NoStop}%
\bibitem [{\citenamefont {Papp}\ \emph {et~al.}(2021)\citenamefont {Papp},
  \citenamefont {Csaba},\ and\ \citenamefont
  {Porod}}]{pappCharacterizationNonlinearSpinwave2021}%
  \BibitemOpen
  \bibfield  {author} {\bibinfo {author} {\bibfnamefont {A.}~\bibnamefont
  {Papp}}, \bibinfo {author} {\bibfnamefont {G.}~\bibnamefont {Csaba}},\ and\
  \bibinfo {author} {\bibfnamefont {W.}~\bibnamefont {Porod}},\ }\bibfield
  {title} {\bibinfo {title} {Characterization of nonlinear spin-wave
  interference by reservoir-computing metrics},\ }\href
  {https://doi.org/10.1063/5.0048982} {\bibfield  {journal} {\bibinfo
  {journal} {Appl. Phys. Lett.}\ }\textbf {\bibinfo {volume} {119}},\ \bibinfo
  {pages} {112403} (\bibinfo {year} {2021})}\BibitemShut {NoStop}%
\bibitem [{\citenamefont {Lee}\ and\ \citenamefont
  {Mochizuki}(2022)}]{leeReservoirComputingSpin2022}%
  \BibitemOpen
  \bibfield  {author} {\bibinfo {author} {\bibfnamefont {M.-K.}\ \bibnamefont
  {Lee}}\ and\ \bibinfo {author} {\bibfnamefont {M.}~\bibnamefont
  {Mochizuki}},\ }\bibfield  {title} {\bibinfo {title} {Reservoir {{Computing}}
  with {{Spin Waves}} in a {{Skyrmion Crystal}}},\ }\href
  {https://doi.org/10.1103/PhysRevApplied.18.014074} {\bibfield  {journal}
  {\bibinfo  {journal} {Phys. Rev. Appl.}\ }\textbf {\bibinfo {volume} {18}},\
  \bibinfo {pages} {014074} (\bibinfo {year} {2022})}\BibitemShut {NoStop}%
\bibitem [{\citenamefont {Nakane}\ \emph {et~al.}(2023)\citenamefont {Nakane},
  \citenamefont {Hirose},\ and\ \citenamefont
  {Tanaka}}]{nakanePerformanceEnhancementSpinWaveBased2023}%
  \BibitemOpen
  \bibfield  {author} {\bibinfo {author} {\bibfnamefont {R.}~\bibnamefont
  {Nakane}}, \bibinfo {author} {\bibfnamefont {A.}~\bibnamefont {Hirose}},\
  and\ \bibinfo {author} {\bibfnamefont {G.}~\bibnamefont {Tanaka}},\
  }\bibfield  {title} {\bibinfo {title} {Performance {{Enhancement}} of a
  {{Spin-Wave-Based Reservoir Computing System Utilizing Different Physical
  Conditions}}},\ }\href {https://doi.org/10.1103/PhysRevApplied.19.034047}
  {\bibfield  {journal} {\bibinfo  {journal} {Phys. Rev. Appl.}\ }\textbf
  {\bibinfo {volume} {19}},\ \bibinfo {pages} {034047} (\bibinfo {year}
  {2023})}\BibitemShut {NoStop}%
\bibitem [{\citenamefont {Namiki}\ \emph {et~al.}(2023)\citenamefont {Namiki},
  \citenamefont {Nishioka}, \citenamefont {Yamaguchi}, \citenamefont
  {Tsuchiya}, \citenamefont {Higuchi},\ and\ \citenamefont
  {Terabe}}]{namikiExperimentalDemonstrationHighPerformance2023}%
  \BibitemOpen
  \bibfield  {author} {\bibinfo {author} {\bibfnamefont {W.}~\bibnamefont
  {Namiki}}, \bibinfo {author} {\bibfnamefont {D.}~\bibnamefont {Nishioka}},
  \bibinfo {author} {\bibfnamefont {Y.}~\bibnamefont {Yamaguchi}}, \bibinfo
  {author} {\bibfnamefont {T.}~\bibnamefont {Tsuchiya}}, \bibinfo {author}
  {\bibfnamefont {T.}~\bibnamefont {Higuchi}},\ and\ \bibinfo {author}
  {\bibfnamefont {K.}~\bibnamefont {Terabe}},\ }\bibfield  {title} {\bibinfo
  {title} {Experimental {{Demonstration}} of {{High}}-{{Performance Physical
  Reservoir Computing}} with {{Nonlinear Interfered Spin Wave
  Multidetection}}},\ }\href {https://doi.org/10.1002/aisy.202300228}
  {\bibfield  {journal} {\bibinfo  {journal} {Advanced Intelligent Systems}\
  }\textbf {\bibinfo {volume} {5}},\ \bibinfo {pages} {2300228} (\bibinfo
  {year} {2023})}\BibitemShut {NoStop}%
\bibitem [{\citenamefont {Nagase}\ \emph {et~al.}(2024)\citenamefont {Nagase},
  \citenamefont {Nezu},\ and\ \citenamefont {Sekiguchi}}]{Nagase2024}%
  \BibitemOpen
  \bibfield  {author} {\bibinfo {author} {\bibfnamefont {S.}~\bibnamefont
  {Nagase}}, \bibinfo {author} {\bibfnamefont {S.}~\bibnamefont {Nezu}},\ and\
  \bibinfo {author} {\bibfnamefont {K.}~\bibnamefont {Sekiguchi}},\ }\bibfield
  {title} {\bibinfo {title} {Spin-wave reservoir chips with short-term memory
  for high-speed estimation of external magnetic fields},\ }\bibfield
  {journal} {\bibinfo  {journal} {Phys. Rev. Appl.}\ }\textbf {\bibinfo
  {volume} {22}},\ \href {https://doi.org/10.1103/physrevapplied.22.024072}
  {10.1103/physrevapplied.22.024072} (\bibinfo {year} {2024})\BibitemShut
  {NoStop}%
\bibitem [{\citenamefont {Namiki}\ \emph {et~al.}(2024)\citenamefont {Namiki},
  \citenamefont {Nishioka}, \citenamefont {Tsuchiya},\ and\ \citenamefont
  {Terabe}}]{namikiFastPhysicalReservoir2024}%
  \BibitemOpen
  \bibfield  {author} {\bibinfo {author} {\bibfnamefont {W.}~\bibnamefont
  {Namiki}}, \bibinfo {author} {\bibfnamefont {D.}~\bibnamefont {Nishioka}},
  \bibinfo {author} {\bibfnamefont {T.}~\bibnamefont {Tsuchiya}},\ and\
  \bibinfo {author} {\bibfnamefont {K.}~\bibnamefont {Terabe}},\ }\bibfield
  {title} {\bibinfo {title} {Fast physical reservoir computing, achieved with
  nonlinear interfered spin waves},\ }\href
  {https://doi.org/10.1088/2634-4386/ad561a} {\bibfield  {journal} {\bibinfo
  {journal} {Neuromorphic Computing and Engineering}\ }\textbf {\bibinfo
  {volume} {4}},\ \bibinfo {pages} {024015} (\bibinfo {year}
  {2024})}\BibitemShut {NoStop}%
\bibitem [{\citenamefont {K{\"o}rber}\ \emph {et~al.}(2023)\citenamefont
  {K{\"o}rber}, \citenamefont {Heins}, \citenamefont {Hula}, \citenamefont
  {Kim}, \citenamefont {Thlang}, \citenamefont {Schultheiss}, \citenamefont
  {Fassbender},\ and\ \citenamefont {Schultheiss}}]{korber2023pattern}%
  \BibitemOpen
  \bibfield  {author} {\bibinfo {author} {\bibfnamefont {L.}~\bibnamefont
  {K{\"o}rber}}, \bibinfo {author} {\bibfnamefont {C.}~\bibnamefont {Heins}},
  \bibinfo {author} {\bibfnamefont {T.}~\bibnamefont {Hula}}, \bibinfo {author}
  {\bibfnamefont {J.-V.}\ \bibnamefont {Kim}}, \bibinfo {author} {\bibfnamefont
  {S.}~\bibnamefont {Thlang}}, \bibinfo {author} {\bibfnamefont
  {H.}~\bibnamefont {Schultheiss}}, \bibinfo {author} {\bibfnamefont
  {J.}~\bibnamefont {Fassbender}},\ and\ \bibinfo {author} {\bibfnamefont
  {K.}~\bibnamefont {Schultheiss}},\ }\bibfield  {title} {\bibinfo {title}
  {Pattern recognition in reciprocal space with a magnon-scattering
  reservoir},\ }\href {https://doi.org/10.1038/s41467-023-39452-y} {\bibfield
  {journal} {\bibinfo  {journal} {Nat. Commun.}\ }\textbf {\bibinfo {volume}
  {14}},\ \bibinfo {pages} {3954} (\bibinfo {year} {2023})}\BibitemShut
  {NoStop}%
\bibitem [{\citenamefont {Schultheiss}\ \emph {et~al.}(2019)\citenamefont
  {Schultheiss}, \citenamefont {Verba}, \citenamefont {Wehrmann}, \citenamefont
  {Wagner}, \citenamefont {K{\"o}rber}, \citenamefont {Hula}, \citenamefont
  {Hache}, \citenamefont {K{\'a}kay}, \citenamefont {Awad}, \citenamefont
  {Tiberkevich}, \citenamefont {Slavin}, \citenamefont {Fassbender},\ and\
  \citenamefont {Schultheiss}}]{schultheiss_excitation_2019}%
  \BibitemOpen
  \bibfield  {author} {\bibinfo {author} {\bibfnamefont {K.}~\bibnamefont
  {Schultheiss}}, \bibinfo {author} {\bibfnamefont {R.}~\bibnamefont {Verba}},
  \bibinfo {author} {\bibfnamefont {F.}~\bibnamefont {Wehrmann}}, \bibinfo
  {author} {\bibfnamefont {K.}~\bibnamefont {Wagner}}, \bibinfo {author}
  {\bibfnamefont {L.}~\bibnamefont {K{\"o}rber}}, \bibinfo {author}
  {\bibfnamefont {T.}~\bibnamefont {Hula}}, \bibinfo {author} {\bibfnamefont
  {T.}~\bibnamefont {Hache}}, \bibinfo {author} {\bibfnamefont
  {A.}~\bibnamefont {K{\'a}kay}}, \bibinfo {author} {\bibfnamefont {A.~A.}\
  \bibnamefont {Awad}}, \bibinfo {author} {\bibfnamefont {V.}~\bibnamefont
  {Tiberkevich}}, \bibinfo {author} {\bibfnamefont {A.~N.}\ \bibnamefont
  {Slavin}}, \bibinfo {author} {\bibfnamefont {J.}~\bibnamefont {Fassbender}},\
  and\ \bibinfo {author} {\bibfnamefont {H.}~\bibnamefont {Schultheiss}},\
  }\bibfield  {title} {\bibinfo {title} {Excitation of {{Whispering Gallery
  Magnons}} in a {{Magnetic Vortex}}},\ }\href
  {https://doi.org/10.1103/PhysRevLett.122.097202} {\bibfield  {journal}
  {\bibinfo  {journal} {Phys. Rev. Lett.}\ }\textbf {\bibinfo {volume} {122}},\
  \bibinfo {pages} {097202} (\bibinfo {year} {2019})}\BibitemShut {NoStop}%
\bibitem [{\citenamefont {K{\"o}rber}\ \emph {et~al.}(2020)\citenamefont
  {K{\"o}rber}, \citenamefont {Schultheiss}, \citenamefont {Hula},
  \citenamefont {Verba}, \citenamefont {Fassbender}, \citenamefont
  {K{\'a}kay},\ and\ \citenamefont
  {Schultheiss}}]{korberNonlocalStimulationThreeMagnon2020}%
  \BibitemOpen
  \bibfield  {author} {\bibinfo {author} {\bibfnamefont {L.}~\bibnamefont
  {K{\"o}rber}}, \bibinfo {author} {\bibfnamefont {K.}~\bibnamefont
  {Schultheiss}}, \bibinfo {author} {\bibfnamefont {T.}~\bibnamefont {Hula}},
  \bibinfo {author} {\bibfnamefont {R.}~\bibnamefont {Verba}}, \bibinfo
  {author} {\bibfnamefont {J.}~\bibnamefont {Fassbender}}, \bibinfo {author}
  {\bibfnamefont {A.}~\bibnamefont {K{\'a}kay}},\ and\ \bibinfo {author}
  {\bibfnamefont {H.}~\bibnamefont {Schultheiss}},\ }\bibfield  {title}
  {\bibinfo {title} {Nonlocal {{Stimulation}} of {{Three-Magnon Splitting}} in
  a {{Magnetic Vortex}}},\ }\href
  {https://doi.org/10.1103/PhysRevLett.125.207203} {\bibfield  {journal}
  {\bibinfo  {journal} {Phys. Rev. Lett.}\ }\textbf {\bibinfo {volume} {125}},\
  \bibinfo {pages} {207203} (\bibinfo {year} {2020})}\BibitemShut {NoStop}%
\bibitem [{\citenamefont {Sebastian}\ \emph {et~al.}(2015)\citenamefont
  {Sebastian}, \citenamefont {Schultheiss}, \citenamefont {Obry}, \citenamefont
  {Hillebrands},\ and\ \citenamefont
  {Schultheiss}}]{sebastianMicrofocusedBrillouinLight2015}%
  \BibitemOpen
  \bibfield  {author} {\bibinfo {author} {\bibfnamefont {T.}~\bibnamefont
  {Sebastian}}, \bibinfo {author} {\bibfnamefont {K.}~\bibnamefont
  {Schultheiss}}, \bibinfo {author} {\bibfnamefont {B.}~\bibnamefont {Obry}},
  \bibinfo {author} {\bibfnamefont {B.}~\bibnamefont {Hillebrands}},\ and\
  \bibinfo {author} {\bibfnamefont {H.}~\bibnamefont {Schultheiss}},\
  }\bibfield  {title} {\bibinfo {title} {Micro-focused brillouin light
  scattering: imaging spin waves at the nanoscale},\ }\bibfield  {journal}
  {\bibinfo  {journal} {Front. Phys.}\ }\textbf {\bibinfo {volume} {3}},\ \href
  {https://doi.org/10.3389/fphy.2015.00035} {10.3389/fphy.2015.00035} (\bibinfo
  {year} {2015})\BibitemShut {NoStop}%
\bibitem [{\citenamefont {Mock}\ \emph {et~al.}(1987)\citenamefont {Mock},
  \citenamefont {Hillebrands},\ and\ \citenamefont
  {Sandercock}}]{mockConstructionPerformanceBrillouin1987}%
  \BibitemOpen
  \bibfield  {author} {\bibinfo {author} {\bibfnamefont {R.}~\bibnamefont
  {Mock}}, \bibinfo {author} {\bibfnamefont {B.}~\bibnamefont {Hillebrands}},\
  and\ \bibinfo {author} {\bibfnamefont {R.}~\bibnamefont {Sandercock}},\
  }\bibfield  {title} {\bibinfo {title} {Construction and performance of a
  {{Brillouin}} scattering set-up using a triple-pass tandem {{Fabry-Perot}}
  interferometer},\ }\href {https://doi.org/10.1088/0022-3735/20/6/017}
  {\bibfield  {journal} {\bibinfo  {journal} {J. Phys. E: Sci. Instr.}\
  }\textbf {\bibinfo {volume} {20}},\ \bibinfo {pages} {656} (\bibinfo {year}
  {1987})}\BibitemShut {NoStop}%
\bibitem [{\citenamefont {Vansteenkiste}\ \emph {et~al.}(2014)\citenamefont
  {Vansteenkiste}, \citenamefont {Leliaert}, \citenamefont {Dvornik},
  \citenamefont {Helsen}, \citenamefont {Garcia-Sanchez},\ and\ \citenamefont
  {Van~Waeyenberge}}]{vansteenkiste_design_2014}%
  \BibitemOpen
  \bibfield  {author} {\bibinfo {author} {\bibfnamefont {A.}~\bibnamefont
  {Vansteenkiste}}, \bibinfo {author} {\bibfnamefont {J.}~\bibnamefont
  {Leliaert}}, \bibinfo {author} {\bibfnamefont {M.}~\bibnamefont {Dvornik}},
  \bibinfo {author} {\bibfnamefont {M.}~\bibnamefont {Helsen}}, \bibinfo
  {author} {\bibfnamefont {F.}~\bibnamefont {Garcia-Sanchez}},\ and\ \bibinfo
  {author} {\bibfnamefont {B.}~\bibnamefont {Van~Waeyenberge}},\ }\bibfield
  {title} {{\selectlanguage {en}\bibinfo {title} {The design and verification
  of {MuMax3}}},\ }\href {https://doi.org/10.1063/1.4899186} {\bibfield
  {journal} {\bibinfo  {journal} {AIP Adv.}\ }\textbf {\bibinfo {volume} {4}},\
  \bibinfo {pages} {107133} (\bibinfo {year} {2014})}\BibitemShut {NoStop}%
\bibitem [{\citenamefont {Heins}\ \emph {et~al.}(2025)\citenamefont {Heins},
  \citenamefont {Kim}, \citenamefont {Körber}, \citenamefont {Faßbender},
  \citenamefont {Schultheiß},\ and\ \citenamefont
  {Schultheiß}}]{heins_christopher_2025_3558}%
  \BibitemOpen
  \bibfield  {author} {\bibinfo {author} {\bibfnamefont {C.}~\bibnamefont
  {Heins}}, \bibinfo {author} {\bibfnamefont {J.-V.}\ \bibnamefont {Kim}},
  \bibinfo {author} {\bibfnamefont {L.}~\bibnamefont {Körber}}, \bibinfo
  {author} {\bibfnamefont {J.}~\bibnamefont {Faßbender}}, \bibinfo {author}
  {\bibfnamefont {H.}~\bibnamefont {Schultheiß}},\ and\ \bibinfo {author}
  {\bibfnamefont {K.}~\bibnamefont {Schultheiß}},\ }\href
  {https://doi.org/10.14278/rodare.3558} {\bibinfo {title} {{Data publication:
  Benchmarking a magnon-scattering reservoir with modal and temporal
  multiplexing}}} (\bibinfo {year} {2025})\BibitemShut {NoStop}%
\bibitem [{\citenamefont {Jaeger}(2001)}]{jaegerShortTermMemory2001}%
  \BibitemOpen
  \bibfield  {author} {\bibinfo {author} {\bibfnamefont {H.}~\bibnamefont
  {Jaeger}},\ }\bibfield  {title} {\bibinfo {title} {Short {{Term Memory}} in
  {{Echo State Networks}}},\ }\bibfield  {journal} {\bibinfo  {journal}
  {GMD-Forschungszentrum Informationstechnik}\ }\href
  {https://doi.org/10.24406/publica-fhg-291107} {10.24406/publica-fhg-291107}
  (\bibinfo {year} {2001})\BibitemShut {NoStop}%
\bibitem [{\citenamefont {Dale}\ \emph {et~al.}(2019)\citenamefont {Dale},
  \citenamefont {Miller}, \citenamefont {Stepney},\ and\ \citenamefont
  {Trefzer}}]{daleSubstrateindependentFrameworkCharacterize2019}%
  \BibitemOpen
  \bibfield  {author} {\bibinfo {author} {\bibfnamefont {M.}~\bibnamefont
  {Dale}}, \bibinfo {author} {\bibfnamefont {J.~F.}\ \bibnamefont {Miller}},
  \bibinfo {author} {\bibfnamefont {S.}~\bibnamefont {Stepney}},\ and\ \bibinfo
  {author} {\bibfnamefont {M.~A.}\ \bibnamefont {Trefzer}},\ }\bibfield
  {title} {\bibinfo {title} {A substrate-independent framework to characterize
  reservoir computers},\ }\href {https://doi.org/10.1098/rspa.2018.0723}
  {\bibfield  {journal} {\bibinfo  {journal} {Proceedings of the Royal Society
  A: Mathematical, Physical and Engineering Sciences}\ }\textbf {\bibinfo
  {volume} {475}},\ \bibinfo {pages} {20180723} (\bibinfo {year}
  {2019})}\BibitemShut {NoStop}%
\end{thebibliography}%

\end{document}